\documentclass[11pt,a4paper]{article}

\usepackage{graphicx,tikz,amssymb,amsmath,times}
\usepackage[affil-it]{authblk}
\usepackage{hyperref}
\hypersetup{
    colorlinks,
    linkcolor={red!10!black},
    citecolor={blue!10!black},
    urlcolor={blue!10!black}
}

\usepackage[utf8]{inputenc}
\usepackage[backend=bibtex,abbreviate=true,sorting=none,sortcites=true,firstinits=true, url=false,doi=false,eprint=false,isbn=false,style=numeric-comp]{biblatex}
\renewbibmacro{in:}{%
  \ifentrytype{article}{}{\printtext{\bibstring{in}\intitlepunct}}}
\newbibmacro{string+doiurlisbn}[1]{%
  \iffieldundef{doi}{%
    \iffieldundef{url}{%
      \iffieldundef{isbn}{%
        \iffieldundef{issn}{%
          #1%
        }{%
          \href{http://books.google.com/books?vid=ISSN\thefield{issn}}{#1}%
        }%
      }{%
        \href{http://books.google.com/books?vid=ISBN\thefield{isbn}}{#1}%
      }%
    }{%
      \href{\thefield{url}}{#1}%
    }%
  }{%
    \href{http://dx.doi.org/\thefield{doi}}{#1}%
  }%
}

\DeclareFieldFormat{title}{\usebibmacro{string+doiurlisbn}{\mkbibemph{#1}}}
\DeclareFieldFormat[article,incollection,inbook,book]{title}%
    {\usebibmacro{string+doiurlisbn}{\mkbibquote{#1}}}
\AtEveryBibitem{
 \clearfield{eprint}
 \clearfield{isbn}
 \clearfield{issn}
 \clearlist{location}
 \clearfield{month}
 \clearfield{note}
}
\providecommand*{\mrm}[1]{\mathrm{#1}}

\newcommand{\cf}{{\it cf.\/}\ }

\newcommand{\ie}{{\it i.e.\/}, }
\newcommand{\eg}{{\it e.g.\/}, }

\newcommand{\set}[1]{\mathrm{#1}}

\newcommand{\RR}{\mathbb{R}}

\newcommand{\LL}{\mathcal{L}}
\newcommand{\KK}{\mathcal{K}}

\newcommand{\ju}{\mathrm{j}}

\newcommand{\Laplace}{\Delta}
\renewcommand{\exp}[1]{\mathrm{e}^{#1}}
\newcommand{\lexp}[1]{\mathrm{e}^{#1}}
\newcommand{\sbj}[1]{\mathop{}\!\j_{#1}}

\newcommand{\RE}{\mathop{\set{Re}}}
\newcommand{\IM}{\mathop{\set{Im}}}

\newcommand{\diff}{\mathop{}\!\mathrm{d}}

\newcommand{\Oh}{\mathcal{O}}

\newcommand{\Div}{\nabla\cdot}
\newcommand{\Rot}{\nabla\times}

\newcommand{\dotp}[2]{\langle #1,#2\rangle}

\newcommand{\eps}{\varepsilon}

\newcommand{\va}{\boldsymbol{a}}

\newcommand{\vb}{\boldsymbol{b}}
\newcommand{\vc}{\boldsymbol{c}}

\newcommand{\vJ}{\boldsymbol{J}}
\newcommand{\vA}{\boldsymbol{A}}

\newcommand{\vF}{\boldsymbol{F}}

\newcommand{\vr}{\boldsymbol{r}}
\newcommand{\vR}{\boldsymbol{R}}

\newcommand{\he}{\hat{\boldsymbol{e}}}

\newcommand{\hz}{\hat{\boldsymbol{z}}}

\newcommand{\hn}{\hat{\boldsymbol{n}}}
\newcommand{\hr}{\hat{\boldsymbol{r}}}
\newcommand{\hth}{\hat{\boldsymbol{\theta}}}
\newcommand{\hphi}{\hat{\boldsymbol{\varphi}}}

\newcommand{\vE}{\boldsymbol{E}}
\newcommand{\vH}{\boldsymbol{H}}

\newcommand{\Wec}{W_{\mrm{e,c}}}
\newcommand{\Wmc}{W_{\mrm{m,c}}}
\newcommand{\Wet}{W_{\mrm{e}}}
\newcommand{\We}{W_{\mrm{e,F}}}
\newcommand{\Wem}{W_{\mrm{em}}}
\newcommand{\Wmt}{W_{\mrm{m}}}
\newcommand{\Wm}{W_{\mrm{m,F}}}
\newcommand{\rP}{P_{\mrm{rad}}}
\newcommand{\tP}{P_{\mrm{tot}}}
\newcommand{\lP}{P_{\mrm{abs}}}

\newcommand{\vN}{\boldsymbol{N}}
\newcommand{\Eo}{\vF_{\mrm{E}}}
\newcommand{\Ho}{\vF_{\mrm{H}}}
\newcommand{\Ao}{\vF_{\mrm{A}}}
\newcommand{\phio}{F_{\phi}}
\newcommand{\Fo}{\vF_{\mrm{N}}}
\newcommand{\psio}{F_\psi}
\newcommand{\hR}{\hat{\boldsymbol{R}}}

\newcommand{\hp}{\hat{\boldsymbol{p}}}
\newcommand{\vp}{\boldsymbol{p}}

\newcommand{\roe}{\rho_{\mrm{e}}}
\newcommand{\rom}{\rho_{\mrm{m}}}

\newcommand{\Rr}{\RR_r^3}

\newcommand{\Qc}{Q}
\newcommand{\QF}{Q_{\mrm{F}}}
\newcommand{\Qe}{Q_{\mrm{e}}}
\newcommand{\Qm}{Q_{\mrm{m}}}
\newcommand{\cP}{P_{\mrm{C}}}

\newcommand{\vJm}{\boldsymbol{J}_{\mrm{m}}}
\newcommand{\vJe}{\boldsymbol{J}_{\mrm{e}}}
\newcommand{\Jmo}{\boldsymbol{J}_{\mrm{m},1}}
\newcommand{\Jeo}{\boldsymbol{J}_{\mrm{e},1}}
\newcommand{\Jmt}{\boldsymbol{J}_{\mrm{m},2}}
\newcommand{\Jet}{\boldsymbol{J}_{\mrm{e},2}}
\newcommand{\JS}{\boldsymbol{\mathcal{J}}}
\newcommand{\br}{{B_{r_0}}}

\newcommand{\gmate}{\mat{\gamma}_{\mrm{e}}}
\newcommand{\gmatm}{\mat{\gamma}_{\mrm{m}}}

\newcommand{\hh}{\hat{\boldsymbol{h}}}
\newcommand{\imp}{\eta}

\newcommand{\Le}{\mathcal{L}_\mrm{e}}
\newcommand{\Lm}{\mathcal{L}_\mrm{m}}
\newcommand{\Lem}{\mathcal{L}_\mrm{em}}

\renewcommand{\vec}[1]{{\boldsymbol#1}}
\newif\ifgreek
\def\testgreek#1{
  \ifx#1\gamma\greektrue\else\ifx#1\Gamma\greektrue\else
  \ifx#1\epsilon\greektrue\else
  \ifx#1\mu\greektrue\else
  \ifx#1\rho\greektrue\else
  \ifx#1\sigma\greektrue\else\ifx#1\Sigma\greektrue\else
     \greekfalse
  \fi\fi\fi\fi\fi\fi\fi}
\newcommand{\mat}[1]{{\testgreek#1\ifgreek\boldsymbol#1\else
                      \mathbf#1\fi}} 

\providecommand*{\mrm}[1]{\mathrm{#1}}

\providecommand*{\diff}{\operatorname{d}\!}
\providecommand*{\ju}{\ensuremath{\mrm{j}}}

\newcommand{\rv}{\vec{r}}
\newcommand{\vd}{\vec{d}}
\newcommand{\rvh}{\hat{\vec{r}}}
\newcommand{\evh}{\hat{\vec{e}}}
\newcommand{\xvh}{\hat{\vec{x}}}

\newcommand{\zvh}{\hat{\vec{z}}}
\newcommand{\Jve}{\vec{J}_{\mrm{e}}}
\newcommand{\Jvm}{\vec{J}_{\mrm{m}}}
\newcommand{\Jm}{\mat{I}}
\newcommand{\Ime}{\mat{I}_{\mrm{e}}}
\newcommand{\Imm}{\mat{I}_{\mrm{m}}}

\newcommand{\herm}{\mrm{H}}
\newcommand{\psiv}{\vec{\psi}}
\newcommand{\qtext}[1]{\quad\text{#1}}
\newcommand{\Km}{\mat{K}}
\newcommand{\Rm}{\mat{R}}
\newcommand{\Fm}{\mat{F}}
\newcommand{\Om}{\mat{0}}
\newcommand{\Xme}{\mat{X}_{\mrm{e}}}
\newcommand{\Xmm}{\mat{X}_{\mrm{m}}}
\newcommand{\Xmee}{\mat{X}_{\mrm{ee}}}
\newcommand{\Xmem}{\mat{X}_{\mrm{em}}}
\newcommand{\Xmme}{\mat{X}_{\mrm{me}}}
\newcommand{\Xmmm}{\mat{X}_{\mrm{mm}}}

\newcommand{\minimize}{\mathop{\mrm{minimize}}}

\newcommand{\subto}{\mrm{subject\ to}}

\newcommand{\Mm}{\mat{M}}

\graphicspath{{./figures/}}

\begin{document}

\title{Stored energies for electric and magnetic current densities}

\author{B. L. G. Jonsson}
\affil{School of Electrical Engineering, 
              KTH Royal Institute of Technology, 
Teknikringen 33, SE-100~40~Stockholm, Sweden}
\author{Mats Gustafsson}
\affil{Electrical and Information Technology, 
           Lund University, 
Box 118, SE-221 00 Lund, Sweden\\
}

\date{\today}

\maketitle

\begin{abstract}
Electric and magnetic current densities are an essential part of electromagnetic theory.  The goal of the present paper is to define and investigate stored energies that are valid for structures that can support both electric and magnetic current densities. Stored energies normalized with the dissipated power give us the Q factor, or antenna Q, for the structure. Lower bounds of the Q factor provide information about the available bandwidth for passive antennas that can be realized in the structure. 
The definition of the stored energies that we propose is valid beyond the leading order small antenna limit. Our starting point is the energy density with subtracted far-field form which we obtain an explicit and numerically attractive current density representation. This representation gives us the insight to propose a coordinate independent stored energy.  
Furthermore, we find here that lower bounds on antenna Q for structures with \eg electric dipole radiation can be formulated as convex optimization problems. We determine lower bounds on both open and closed surfaces that support electric and magnetic current densities. 

The here derived representation of stored energies has in its electrical small limit an associated Q factor that agrees with known small antenna bounds. These stored energies have similarities to earlier efforts to define stored energies.  However, one of the advantages with this method is the above mentioned formulation as convex optimization problems, which makes it easy to predict lower bounds for antennas of arbitrary shapes. The present formulation also gives us insight into the components that contribute to Chu's lower bound for spherical shapes.  We utilize scalar and vector potentials to obtain a compact direct derivation of these stored energies. Examples and comparisons end the paper.

\end{abstract}

\section{Introduction}

The antenna Q, or the Q-factor, for an antenna is a dimensionless number that provides information about the impedance bandwidth about the device~\cite{Hansen1981,Volakis+etal2010,Yaghjian+Best2005,Gustafsson+etal2015b}. Antenna Q is proportional to the ratio of stored energy to the dissipated energy per cycle. For antennas that operate at a single resonance in their input impedance, it provides an accurate description of the impedance bandwidth. 
In the present paper, we investigate how to define electric and magnetic stored energies for somewhat larger structures, such that the electrically small limit agrees with the classical results~\cite{Chu1948,Hansen+Collin2009,Thal2006,Thal2009,Gustafsson+Jonsson2015b,Jonsson+Gustafsson2015}. We focus on a stored energy definition that is based on a current density representation, which is coordinate independent and rather straight forward to determine. The definition still suffers from the fact that as the size of the object approaches a wavelength it can become very small (zero) for certain wavelength regions.  The definition is for arbitrarily shaped regions including open and closed surfaces, which support both electric and magnetic current densities. A-priori information about the lowest possible antenna Q and the largest partially realized gain over antenna Q-ratio, $G/Q$, for a given volume are important in electromagnetic design~\cite{Hansen1981,Volakis+etal2010,Yaghjian+Best2005,Gustafsson+etal2015b}. Such information can provide guidelines to specify the minimal required size of a device. 
The results in this paper provide a link to a partial understanding on how antenna Q is related to the shape of the current density support. We illustrate the shape-dependence of antenna Q at the end of the paper, and note that the small electric limit (leading order) of these energies are illustrated in~\cite{Jonsson+Gustafsson2015}. This understanding is naturally implicit, and it is given as an optimization problem for arbitrarily shaped objects. For the small electric size this optimization problem can be solved explicitly for arbitrary shapes~\cite{Yaghjian+etal2013,Jonsson+Gustafsson2013a,Jonsson+Gustafsson2015} and when these explicit solutions are compared with the sphere~\cite{Chu1948,Thal2006} they capture the leading order term. The here defined stored energies aim to obtain all terms.  

Different approaches to optimize currents to find optimal  energies, directivity or powers transfer have been considered in~\eg \cite{Wheeler1965,Margetis+etal1998,Geyi2003,Yaghjian+Stuart2010,Vandenbosch2010,Vandenbosch2011,Gustafsson+etal2012a,Gustafsson+Nordebo2013,Razavi+etal2015,Gustafsson+Jonsson2015a,Jonsson+Gustafsson2015,Jelinek+Capek2016,Gustafsson+etal2016a}.
Another advantage of the here proposed stored energy is that it can be combined with constraints on far-fields such as \eg a directivity constraint as to form a convex optimization problem~\cite{Gustafsson+Nordebo2013,Gustafsson+etal2016a}.
  
In the literature there are at least four different approaches to determine bounds on the antenna Q and implicitly on the bandwidth~\cite{Gustafsson+etal2015b}. The first approach is based on circuit models~\cite{Wheeler1947,Chu1948,Thal2006,Thal2009}. The second approach is based on defining a concept of stored energies for vector modes of a canonical shape. This approach includes the work on spherical shapes~\cite{Chu1948, Harrington1960,Collin+Rothschild1964,Hansen1981,McLean1996}, cylindrical~\cite{ Collin+Rothschild1964}, and spheroidal \cite{Foltz+Mclean1999,Sten+etal2001} shapes using vector-modes.   
The third method to derive bounds is based on sum rules, where the antenna is modeled as a passive system~\cite{Gustafsson+etal2007a,Gustafsson+etal2009a,Sohl+Gustafsson2008a}. This approach provides bounds for arbitrarily shaped antennas and is not limited to electrically small objects. 
Beyond antenna Q, sum-rules can also give limitations for array antennas, see e.g.
~\cite{Jonsson+Gustafsson2010,Doane+etal2013,Jonsson+etal2013,Jonsson2014a}. The fourth method is to determine antenna Q-bounds directly from minimization of the stored energies for all current densities~\cite{Yaghjian+Stuart2010,Vandenbosch2010,Vandenbosch2011, Gustafsson+etal2012a,Gustafsson+Nordebo2013, Jonsson+Gustafsson2013a,Yaghjian+etal2013,Geyi2015,Jonsson+Gustafsson2015},
 see also~\cite{Gustafsson+etal2015b} for an overview of the methods and~\cite{Best2015,Gustafsson+etal2010a,Cismasu+Gustafsson2014} for comparisons between the bounds and antenna performance. The present work is in this fourth category. The methods 1, 2 and 4 that use stored energy are somewhat similar to each other.  The fourth approach to stored energy provides an attractive method to evaluate the stored energies as it is explicit in the current density. This has already lead to several application and approaches to explicitly evaluate and investigate both antenna Q and gain over Q for theoretical and practical cases~\cite{Gustafsson+etal2012a,Gustafsson+Nordebo2013,Cismasu+Gustafsson2014,Gustafsson+etal2016a}.

Beyond these methods to determine bounds of antenna Q, there are several approaches to calculate antenna Q for a given structure, including methods based on the frequency derivatives of the input impedance~\cite{Yaghjian+Best2005,Capek+etal2012,Capek+etal2013} or through Brune circuit synthesis~\cite{Gustafsson+Jonsson2015a}. These methods essentially differ by the use of local respective global spectrum information of the input impedance, see also~\cite{Capek+etal2016} for a time-domain approach. One step in the validation process of our definition of stored energies is to study the sphere. Chu's bound for antennas circumscribed by a sphere requires both electric and magnetic current densities. A further advantage in using the sphere in the validation process is that it is thoroughly analyzed, studied and discussed in the literature, see \eg~\cite{Sievenpiper+etal2012}. It is hence a good benchmark to validate our expression against since it also includes higher order contributions beyond the leading order electrically small term.  One nice feature of all these different theories is that when they are compared for electrically small objects radiating as an electric dipole, the differences between them are rather small~\cite{Gustafsson+Jonsson2015b,Gustafsson+Jonsson2015a}. The here defined stored energies also agree with the electrically small limit with the polarization bounds for the Q-factor~\cite{Gustafsson+etal2012a,Yaghjian+etal2013,Jonsson+Gustafsson2015}. Extensions to antennas with losses and embedded in lossy media have been considered in~\cite{Gustafsson+etal2014,Yaghjian+etal2013,Gustafsson+etal2015a}.

The here derived results for stored energies are related to a symmetric generalization of stored energy proposed by Vandenbosch~\cite{Vandenbosch2010}, see also~\cite{Vandenbosch2013a,Vandenbosch2013b}. We follow the approach of~\cite{Gustafsson+etal2012a,Gustafsson+Jonsson2015b}, here extended to include both electric and magnetic current densities. Our approach is also similar to~\cite{Kim2016} which also include both electric and magnetic currents. The results in~\cite{Kim2016} are based on the assumption that the magnetic currents can be determined from the electric currents under the condition that the field vanishes within the volume. This is an accurate approximation for many geometries and currents, however note that that minimizing electric current is non-unique~\cite{Gustafsson+etal2012a}.
Our approach does not require any such assumptions, nor is it limited to closed volumes. The approaches are similar for the cases when the coupling terms between the electric and magnetic currents in the optimization problem considered in this paper can be neglected.
Note that the results of~\cite{Chu1948,Harrington1960,Yaghjian+etal2013,Jonsson+Gustafsson2015,Kim2016} utilize waves that are generated by both electric and magnetic current sources. 
In the present derivation we show that the electric and magnetic sources enter in the stored energies in a symmetric and explicit way, and that the terms are straight forward to calculate using the well-known electric field integral equation (EFIE) kernels and the magnetic field integral equation (MFIE) kernels.

The paper consists of ten sections, including this introduction. In Sec.~\ref{Def} we start from Maxwell's equations, and define the main quantities in terms of the fields and the far fields, in particular we define the far-field subtracted stored energies.
In Sec.~\ref{sec:Stored}, we derive the expressions for the far-field subtracted stored energies in terms of the electromagnetic sources, and use this to define stored energies. The radiation intensity and radiated power is similarly expressed in terms of the sources in Sec.~\ref{sec:Power}. The electrically small case is briefly discussed in Sec.~\ref{sec:small}. We study the coordinate dependence of the different terms of far-field subtracted stored energy in Sec.~\ref{Coord}. In Sec.~\ref{sec:Matrix}, we provide a bi-linear matrix representation of the stored energies that is suitable to implement numerically. In Sec.~\ref{sec:opt}, we formulate the convex optimization problems. We determine the Q factor for some families of shapes in Sec.~\ref{sec:Ill}.  The paper ends with a conclusion and an appendix. 

\section{Definition of far-field subtracted stored energies and the radiated power}\label{Def}

\begin{figure}
\centering
\begin{tikzpicture}[scale=1,very thick]
\draw[fill,color=blue!20,samples=200,smooth] plot (canvas polar
cs:angle=\x r,radius={54-14*cos(3*\x r)});
\draw[dashed,samples=200,smooth] plot (canvas polar
cs:angle=\x r,radius={1.2 r});
\draw (-.2,-.2) node {$\vJe,\vJm,\roe,\rom$};
\draw[-latex] (0,0) -- (1.72,1.72) node [midway,above=2mm] {$a$};
\draw (.7,-1.4) node {$V$} (-.8,-1.6) node {$\partial V$};
\draw [-latex] (0,0)  ++(20:2.4) -- +(20:1.2) node[right] {$\hr$};
\draw (-1.75,1.75) node[left] {$V_a$};
\end{tikzpicture}
\caption{The figure illustrates the support, $V$, of the sources. This region has a boundary $\partial V$. The region $V$ is bounded and hence it can be enclosed within a sphere $V_a$ of radius $a$.}\label{support}
\end{figure}
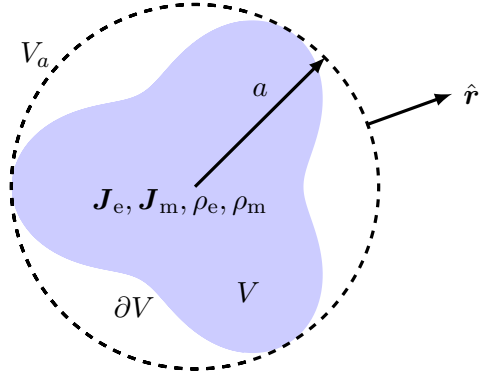

Starting from Maxwell's equations, we here introduce the basic quantities of radiated power, $\rP$, and the far-field subtracted stored electric and magnetic energies, $\We$, and, $\Wm$, respectively. Towards this end, let $V$ be a bounded domain in $\RR^3$ corresponding to the joint support of the electric and magnetic current- and charge-densities $\vJe,\roe,\vJm,\rom$, see Fig.~\ref{support}.  Note that the upcoming analysis makes rather few assumptions on $V$. For example it can be a surface with the usual change of $\vJ\diff V\rightarrow \vJ_s \diff S$, where $\vJ_s$ is a surface current density.  The final restrictions on $V$ are implicit. We require that $V$ is regular enough such that the quadratic form of the far-field subtracted stored energies can be defined, for a short discussion, see Sec.~\ref{sec:3.2}.

Maxwell's equations in free space with electric and magnetic current and charge densities are given by:
\begin{align}\label{MEa}
\nabla \times \vE + \ju \imp k \vH & = - \vJm, && 
\nabla \cdot \vE  = \frac{\roe}{\eps} = \frac{-\imp}{\ju k}\nabla\cdot \vJe, \\
\nabla \times \vH - \frac{\ju k}{\imp}\vE &  = \vJe , &&
\nabla\cdot \vH = \frac{\rom}{\mu} = \frac{-1}{\ju k \imp}\nabla\cdot \vJm,\label{MEb}
\end{align}
To solve them, we also need the Silver-M\"uller radiation condition~\cite{Muller1948,Silver1949,Wilcox1956}.
In this paper we let $\eps$, $\mu$ and $\imp=\sqrt{\mu/\eps}$ be the free space permittivity, permeability and impedance, respectively. The electric and magnetic fields are denoted $\vE$ and $\vH$, the dispersion relation between the wavenumber, $k$, and the angular frequency, $\omega$, is $k=\omega\sqrt{\eps\mu}$. 
For notational convenience we switch back and forth between $\omega$ and $k$. 
In representing Maxwell's equations above we use the time-convention $\lexp{\ju\omega t}$, where $t$ is time.  

Poynting's theorem in its differential form for electromagnetic fields generated by both electric and magnetic currents reads:
\begin{equation}\label{PT}
-\Div(\vE \times \vH^*) + \ju\omega(\eps |\vE|^2-\mu |\vH|^2) = \vE\cdot \vJe^* + \vH^*\cdot \vJm,
\end{equation}
where $*$ indicates the complex conjugate. We can hence find the complex power as 
\begin{equation}\label{cP}
\cP=-\frac{1}{2}\int_V \vE\cdot \vJe^*+\vH^*\cdot\vJm\diff V,
\end{equation}
and the associated radiated power given as the real-valued part of the complex power, 
\cf~\cite{Jackson1999}. We find
\begin{equation}\label{rad}
\rP=\RE\cP=\frac{1}{2\imp}\int_\Omega |\Eo|^2\diff \Omega=\frac{1}{2}\RE \int_{\partial V_a} \vE\times \vH^* \cdot \hr\diff S = -\frac{1}{2}\RE\int_V \vE\cdot \vJe^* + \vH^*\cdot \vJm \diff V,
\end{equation}
where $\Omega$ is the unit sphere and $\diff \Omega$ is the surface element on the unit sphere. Here, $\Eo$ is the electric far-field amplitude \ie $\vE \rightarrow \Eo \frac{\lexp{-\ju k r}}{r}$ as $r\rightarrow \infty$. The sphere of radius $a$ is denoted $V_a$ with boundary $\partial V_a$, see Fig.~\ref{support}, we use $\diff S$ and $\diff V$ to denoted surface and volume element respectively. 
Let $\vr\in\RR^3$ be a vector with length $r=|\vr|$ and the associated unit-vector is $\hr=\vr/r$. Note that the spatial dependence of the fields is  suppressed in this paper. The identity~\eqref{rad} is derived under the implicit assumption that $V$ is bounded by utilizing that the radiated power through the circumscribing surface $\partial V_a$ is conserved.

It is well known that the total electric and magnetic energies associated with~\eqref{MEa} and~\eqref{MEb} are unbounded in $\RR^3$ since the energy densities decay as $\frac{1}{r^2}$, as $r\rightarrow \infty$.  The classical approach to solve this problem is to subtract the far-field energy density, see \eg~\cite{Collin+Rothschild1964,Geyi2003,Yaghjian+Best2005,Fante1969,Gustafsson+Jonsson2015b,Gustafsson+Jonsson2015a} to obtain finite stored energies:
\begin{equation}\label{WEM}
\We = \frac{\eps}{4}\int_{\Rr} |\vE|^2 - \frac{|\Eo|^2}{r^2} \diff V
\qtext{and}\ \ \
\Wm = \frac{\mu}{4}\int_{\Rr} |\vH|^2 - \frac{|\Eo|^2}{\eta^2 r^2} \diff V,
\end{equation}
where $\int_{\Rr}$ is a short hand notation for $\lim_{r_0\rightarrow \infty} \int_{|\vr|\leq r_0}$, see Appendix~\ref{A}. These energies $\We$ and $\Wm$ we denote \emph{far-field subtracted stored electric and magnetic energies}, respectively.  However, there are also other possible decomposition methods to defined finite stored energies, and the merit of a particular choice of decomposition needs to be evaluated on its usefulness.  These far-field stored subtracted energies~\eqref{WEM} are thoroughly investigated for {\it electric} sources $(\roe,\vJe)$ in~\cite{Gustafsson+etal2012a,Gustafsson+Jonsson2015b,Gustafsson+Jonsson2015a}.
Furthermore, it was shown that for electric sources $\We$, $\Wm$ agree with the stored energies~\cite{Vandenbosch2010} proposed by Vandenbosch for many cases. 
There are thus several possible stored energies, with associated Q factors.
A way to interpret these different energies is to think of them as approximations of true stored energies within a small interval that gives a $ka$-uncertainty of the Q factor~\cite{Gustafsson+Jonsson2015b}. 

The definition~\eqref{WEM} of the far-field subtracted stored energies is attractive in several ways. It removes the far field in such a way that these stored energies can be defined for arbitrarily shaped objects.  Their leading order term give rise to bounds for the Q factor in terms of the electric and magnetic polarizabilities~\cite{Gustafsson+etal2012a,Gustafsson+Jonsson2015b,Gustafsson+Jonsson2015a,Yaghjian+etal2013}.  However it is also known that this expression is coordinate dependent for larger structures and that it can be negative for certain frequencies and sources~\cite{Gustafsson+etal2012a,Yaghjian+Best2005}. In the present paper, we define coordinate independent stored electric $\Wet$ and magnetic $\Wmt$ energies based on  the far-field subtracted stored energies $\We$ and $\Wm$, respectively. The definition is given in Sec.~\ref{sec:Stored} where $\We$ and $\Wm$ are thoroughly analyzed. 

One of the interesting quantities to study associated with a stored energy is the Q-factor for antennas, the antenna Q, let $Q=\max(\Qe,\Qm)$ with
\begin{equation}
\Qe=\frac{2\omega \Wet}{\tP}
\qtext{and } \Qm=\frac{2\omega \Wmt}{\tP}.
\label{eq:Qdef}
\end{equation}
Here $\tP=\rP+\lP$ denotes the sum of radiated power and ohmic heating. 
Similarly upon defining the partial gain of the antenna as~\cite{IEEE1993}
\begin{equation}
G(\hr,\he) =  4\pi \frac{P(\hr,\he)}{\tP},
\end{equation}
where $P(\hr,\he)$ is the partial radiation intensity in the direction $\hr$ with polarization $\he$, see \eqref{UPa} and~\eqref{UP}.
We note that the partial gain Q-factor quotient $G/Q$ becomes:
\begin{equation}
\frac{G(\hr,\he)}{Q} = \frac{2\pi P(\hr,\he)}{\omega \max\{\Wet,\Wmt\}}.
\label{eq:GoQ}
\end{equation}


\section{The source term representation of the far-field stored energies}\label{sec:Stored}

The goal in this section is to derive integral expressions for the far-field subtracted stored energies~\eqref{WEM} in terms of the current densities that are both explicit and numerically accessible.  
There are at least two different approaches towards this end.  
One method is pursued in Vandenbosch~\cite{Vandenbosch2010} for electric currents.  It is based on a fundamental solution approach applied directly to the fields, see also~\cite{Vandenbosch2013a,Vandenbosch2013b} for a derivation in the time domain.  
We find that it is easier to approach this derivation via the potentials~\cite{Gustafsson+Jonsson2015b,Jonsson+Gustafsson2015}.  
The approach followed below first express the far-field subtracted stored energies in terms of the potentials. This representation allows us to identify three kernels that need to be calculated to obtain the desired source representation of~\eqref{WEM}.  From this representation it is clear that \eqref{WEM} has a weak coordinate dependence which has similarities to \cite{Yaghjian+Best2005}.  The source representation of~\eqref{WEM} consists of a core part that is inherently coordinate independent and a part that can have a coordinate dependence. The latter part is small when $ka$ is small. The core part is here used to define the coordinate independent stored eneriges, at the end of the section.

For electric sources, it is observed that the leading order far-field subtracted stored energies give rise to Q factors with a lower bound expressed in terms of polarizabilities.  These bounds provide an upper bound for the impedance bandwidth within the shape in the leading order of small $ka$. Well-designed antennas with electric sources have been shown to reach the bound~\cite{Gustafsson+etal2012a}. The antenna Q bounds for small antennas with both electric and magnetic sources are determined in~\cite{Jonsson+Gustafsson2015}. We are here interested in how to define stored energies beyond the leading order case. 

\subsection{Potential formulation}

To set notations and to introduce the method, we start by expressing the far-field subtracted stored electrical energy $\We$ in terms of the current densities. We approach this derivation through the potentials $\phi,\vA,\psi$, and $\vN$, corresponding to the electric scalar potential, magnetic vector potential, magnetic scalar potential, and the electric vector potential, respectively. We find the electric and magnetic fields as:
\begin{equation}\label{pp}
\vE = -\nabla \phi -  \ju\omega \vA - \frac{1}{\eps} \Rot\vN
\qtext{and}\ \ \ \
\vH = -\nabla \psi - \ju \omega \vN + \frac{1}{\mu} \Rot \vA,
\end{equation}
respectively.
These potentials are not unique, and we select the Lorenz gauge: $\Div\vN + \ju\omega \eps\mu\psi=0$ and $\Div\vA + \ju\omega \eps\mu\phi=0$. With this choice of gauge all potentials satisfy a non-homogeneous scalar or vector Helmholtz equation: $\mathcal{H}\vA=-\mu\vJe$, $\mathcal{H}\phi=-\rho/\eps$, $\mathcal{H}\psi=-\rom/\mu$ and $\mathcal{H}\vN=-\eps\vJm$, where $\mathcal{H}=\Laplace+k^2$. We thus have the potentials expressed in terms of the sources as
\begin{align}\label{AG}
\vA(\vr) = \frac{\mu}{4\pi}\int_V \vJe(\vr_1)\frac{\lexp{-\ju k |\vr-\vr_1|}}{|\vr-\vr_1|}\diff V_1, &&\ 
\phi(\vr) = \frac{1}{4\pi\eps}\int_V \roe(\vr_1)\frac{\lexp{-\ju k |\vr-\vr_1|}}{|\vr-\vr_1|}\diff V_1, \\ \label{FG}
\vN(\vr) = \frac{\eps}{4\pi}\int_V \vJm(\vr_1)\frac{\lexp{-\ju k |\vr-\vr_1|}}{|\vr-\vr_1|}\diff V_1, &&\ 
\psi(\vr) = \frac{1}{4\pi\mu}\int_V \rom(\vr_1)\frac{\lexp{-\ju k |\vr-\vr_1|}}{|\vr-\vr_1|}\diff V_1. 
\end{align}

Denote the far-field amplitude of $\vA$ with $\Ao$, where $\vA\rightarrow \Ao\frac{\lexp{-\ju k r}}{r}$ as $r\rightarrow \infty$ and similarly for the far-field amplitude of each of the potentials. The free-space Green's function representation of the potentials~\eqref{AG}--\eqref{FG} gives us directly far-field amplitudes as 
\begin{align}\label{Ao}
\Ao(\hr) = \frac{\mu}{4\pi} \int_V \lexp{\ju k \hr\cdot \vr_1} \vJe(\vr_1)\diff V_1, &&\
\phio(\hr) = \frac{1}{4\pi\eps} \int_V \lexp{\ju k \hr\cdot \vr_1}\roe(\vr_1)\diff V_1,\\
\Fo(\hr) = \frac{\eps}{4\pi} \int_V \lexp{\ju k \hr\cdot \vr_1} \vJm(\vr_1)\diff V_1, &&
\
\psio(\hr) = \frac{1}{4\pi\mu} \int_V \lexp{\ju k \hr\cdot \vr_1}\rom(\vr_1) \diff V_1.\label{psio}
\end{align}
The expression of the electric field in the potentials~\eqref{pp} supply the relation between the far-field amplitude of the electric field, $\Eo$ and the far-field amplitudes of the potentials through: 
\begin{equation}
\Eo= -\ju \omega \Ao + \ju k \hr \phio + \frac{1}{\eps}\ju k \hr \times \Fo.
\end{equation}

Recall that the electric far field is orthogonal to its propagation direction for a source of bounded extent, \ie $\hr\cdot \Eo=0$. This relation yields that $\omega \hr\cdot\Ao = k\phio$, and similarly $\hr\cdot\Ho=0$ yields $\omega \hr\cdot \Fo = k\psio$, we find:
\begin{equation}\label{Eo}
|\Eo|^2 
= \omega^2 |\Ao|^2 - k^2 |\phio|^2 + \frac{k^2}{\eps^2} |\Fo|^2 - k^2\imp^2|\psio|^2  - 2\frac{\omega k }{\eps}\RE \{\Ao\cdot \hr\times \Fo^* \}.
\end{equation} 
We thus have the first term in~\eqref{WEM}. We also express $|\vE|^2$ in terms of the potentials using~\eqref{pp}:
\begin{multline}\label{vEq}
|\vE|^2 
 = \omega^2 |\vA|^2 + |\nabla \phi|^2 - \frac{1}{\eps^2}\RE\{\vN\cdot \Laplace\vN^*\} - k^2\imp^2|\psi|^2 + \frac{1}{\eps^2}\RE\big\{\Div\big(\vN\Div\vN^* - (\Rot\vN^*)\times\vN\big)\big\} \\
- 2k^2|\phi|^2 +2\RE\{\ju \omega \Div( \vA \phi^*)\} + \frac{2}{\eps}\RE\{\ju \omega \vA\cdot \nabla\times \vN^*\} + \frac{2}{\eps}\RE\{ \nabla\cdot (\phi\Rot\vN^*)\} .
\end{multline}
To obtain~\eqref{vEq} we used the standard vector identities $\vb\cdot\nabla\phi = \Div(\phi\vb) - \phi\Div\vb$ with $\vb=\vA^*$ and $\vb=\Rot\vN^*$ respectively. We also used the identity
\begin{equation}
|\Rot \vN|^2 = -\RE \{\vN\cdot\Laplace\vN^*\} - |\Div\vN|^2 +  \RE \Div\big(\vN\Div\vN^* - (\Rot\vN)^*\times \vN\big).
\end{equation}

All divergence terms in~\eqref{vEq} vanish upon integrating them over the domain $\Rr$. To show this fact, we investigate each divergence term below. The volume integral of terms of the form $\Div\vb$ together with Gauss theorem yield that it is sufficient to show that $\hr\cdot \vb$ asymptotically vanishes as $r^{-3}$ or faster as $r\rightarrow \infty$. This follows since we study terms of the type $\lim_{r\rightarrow \infty} r^2\RE\int_{\Omega} \hr\cdot \vb \diff\Omega$, where $\Omega$ is the unit sphere.
First consider the $\vN$-term:
\begin{equation}
(\hr\cdot \vN)\Div\vN^* - \hr\cdot\big((\Rot\vN^*)\times \vN\big) \rightarrow 
\frac{\ju k |\Fo|^2}{r^2} + \Oh(r^{-3})
\qtext{as }
r\rightarrow \infty,
\end{equation}
where we have used the asymptotic behavior of $\vN$. Here $\Oh(r^{-3})$ is the order (big O) of the rest-term, using the Bachmann-Landau notation as $r\rightarrow \infty$. The real valued part of this integrand vanishes with the rate $r^{-3}$ as $r\rightarrow \infty$.

For the second divergence term in~\eqref{vEq}, recall $\hr\cdot \vE = \Oh(r^{-2})$ as $r\rightarrow \infty$. Thus asymptotically we find that  
\begin{equation}
\ju \omega \hr\cdot \vA\phi^* \rightarrow \frac{\ju k|\phio|^2}{r^2} + \Oh(r^{-3}) ,
\end{equation}
Clearly also this term vanishes upon taking the real part with the rate $r^{-3}$, for large enough radius. For the last divergence term we find with increasing radius that 
\begin{equation}
\phi \hr \cdot \Rot \vN^* \rightarrow \frac{\ju k \phio\hr\cdot (\hr\times \Fo^*)}{r^2} +\Oh(r^{-3})= \Oh(r^{-3}).
\end{equation}
We have hence showed that all divergence terms vanish at a rate $\Oh(r^{-3})$ as $r\rightarrow \infty$, and in the limit these terms will not contribute to the far-field subtracted stored energies~\eqref{pW}. 
Note that in the above derivation, we use the spherical shape of the integral domain in the limit process since we have used the surface normal $\hr$ to show these identities.

We insert the energy density~\eqref{vEq} and far-field amplitudes \eqref{Eo} into the far-field subtracted stored electrical energy~\eqref{WEM} to find:
\begin{multline}\label{pW}
\We = \frac{\eps}{4}\int_{\Rr} |\vE|^2-\frac{|\Eo|^2}{r^2}\diff V  
 = \frac{\eps}{4}\int_{\Rr}  
(|\nabla \phi|^2 -k^2|\phi|^2)
- \frac{1}{\eps^2}( \RE\{\vN\cdot \Laplace\vN^*\} +k^2|\vN|^2) 
\\
+ \frac{k^2}{\eps^2}(|\vN|^2 - \frac{|\Fo|^2}{r^2})
 +\omega^2 (|\vA|^2 -\frac{|\Ao|^2}{r^2})  
 - k^2 (|\phi|^2 - \frac{|\phio|^2}{r^2})\\ 
-k^2\imp^2( |\psi|^2 -\frac{|\psio|^2}{r^2}) 
+ \frac{2\omega }{\eps}\RE\Big\{\ju \vA\cdot \nabla\times \vN^* 
-\frac{\ju \Ao \cdot \ju k \hr\times \Fo^*}{r^2}\Big\} \diff V.
\end{multline}
The terms are collected in pairs. Note that the first two pairs are closely related to Helmholtz equation, the next four pairs have a mutually similar structure, and then we have a cross term. We thus have three kinds of terms which will be reduced to their source representation below.

\subsection{Source term representation of the far-field subtracted stored electrical energy}\label{sec:3.2}

To start the process to express~\eqref{pW} in terms of sources, we first note that the integrand $|\nabla\phi|^2-k^2|\phi|^2$ can be rewritten into a Helmholtz operator plus correction terms using the identity $|\nabla \phi|^2 = \Div(\phi\nabla\phi^*)-\phi\Laplace\phi^*$. This, together with the continuity equation gives the identity
\begin{multline}\label{L1}
\int_{\Rr} |\nabla\phi|^2 - k^2|\phi|^2 \diff V = \frac{1}{\eps}\RE \int_V \phi\roe^*\diff V  \\
=\frac{\imp^2}{k^2} \int_V\int_V \big(\nabla_1\cdot\vJe(\vr_1)\big)\frac{\cos(k|\vr_1-\vr_2|)}{ 4\pi|\vr_1-\vr_2|}\nabla_2\cdot\vJe^*(\vr_2)\diff V_1\diff V_2 ,
\end{multline}
which is the desired source representation of the first term in~\eqref{pW}. Similarly the corresponding identity for vectors follows as:
\begin{multline}\label{L2}
-\frac{1}{\eps^2}\RE \int_{\Rr} \vN\cdot \Laplace \vN^* + k^2|\vN|^2 \diff V =  \frac{1}{\eps}\RE \int_V \vN\cdot \vJm^*\diff V   \\ = \int_V\int_V \Jmo\cdot\Jmt^*\frac{\cos(kR_{12})}{ 4\pi R_{12}}\diff V_1\diff V_2.
\end{multline}
Here we use the notation $\vR_{12}=\vr_1-\vr_2$, $R_{12}=|\vR_{12}|$, $\Jmo=\vJm(\vr_1)$, and similar for $\Jmt$ and later with $\Jeo,\Jet$.  
Both of these expressions, \eqref{L1} and \eqref{L2}, are symmetric real-valued quadratic forms with respect to either $\vJe$ or $\vJm$ and both can be extracted from the EFIE-representation of Maxwell's equations. 

The next term in~\eqref{pW} have the structure, $|\vN|^2-|\Fo|^2/r^2$. Replacing the magnetic vector potential $\vN$ by its Green's function representation~\eqref{FG} we find 
\begin{equation}\label{dF}
\frac{k^2}{\eps^2}\int_{\Rr} |\vN|^2-\frac{|\Fo|^2}{r^2} \diff V =   
k^2 \RE \int_V\int_Vg(\vr_1,\vr_2)  \vJm(\vr_1)\cdot\vJm^*(\vr_2) \diff V_1\diff V_2,
\end{equation}
where $g$ is the kernel given by:
\begin{equation}\label{g}
g(\vr_1,\vr_2) = \int_{\Rr} G_1G_2^*- \frac{\lexp{\ju k \hr \cdot (\vr_1-\vr_2)}}{(4\pi r)^2} \diff V =
-\frac{\sin(kR_{12})}{8 k \pi} -  \ju \frac{r_1^2-r_2^2}{8\pi R_{12}}\sbj{1}(kR_{12}), 
\end{equation}
see~\eqref{ggg} in Appendix~\ref{A} for its derivation. Let the free space Green's function (fundamental solution) corresponding to Helmhotlz equation be denoted by $G(\vr)=\frac{\lexp{-\ju k |\vr|}}{4\pi|\vr|}$. We also use the notation $G_n=G(\vr-\vr_n)$, $n=1,2$ and $G_{12}=G(\vr_1-\vr_2)$ below.  Here $\sbj{n}$ is the spherical Bessel function of order $n$.

The next three terms all have an identical integral kernel, and it is straight forward to show that:
\begin{align}
 \omega^2\int_{\Rr} |\vA|^2-\frac{|\Ao|^2}{r^2}\diff V &= \imp^2 k^2 \RE\int_V\int_V g(\vr_1,\vr_2)\vJe(\vr_1)\cdot\vJe^*(\vr_2)\diff V_1\diff V_2,\label{dA} \\
k^2\int_{\Rr} |\phi|^2 - \frac{|\phio|^2}{r^2}\diff V &= \eta^2
\RE \int_V\int_V g(\vr_1,\vr_2)\big(\nabla_1\cdot\vJe(\vr_1)\big)\big(\nabla_2\cdot\vJm^*(\vr_2)\big) \diff V_1\diff V_2,\\
k^2\imp^2 \int_{\Rr} |\psi|^2 -\frac{|\psio|^2}{r^2}\diff V & = \RE \int_V\int_V g(\vr_1,\vr_2)\big(\nabla_1\cdot\vJm(\vr_1)\big)\big(\nabla_2\cdot\vJm^*(\vr_2)\big)\diff V_1\diff V_2.\label{dp}
\end{align}
We have thus described the two first kind of terms in~\eqref{pW}. Below we show that the last term in~\eqref{pW}, the cross term, is a gradient of the kernel $g$. This can be seen by once again substituting the Green's function for the potentials, to obtain 
\begin{equation}
\frac{2\omega}{\eps} \RE \int_{\Rr} \ju \vA\cdot \Rot\vN^* - \frac{\ju \Ao\cdot \ju k \hr\times \Fo^*}{r^2}\diff V 
= 
2 k \imp  \IM  \int_V \int_V
 (\Jmt^*\times \Jeo)\cdot   \nabla_2  g(\vr_1,\vr_2)
\diff V_1\diff V_2. 
\end{equation}
The kernel $\nabla_2 g$ also have an explicit representation, see~\eqref{nggg} in Appendix~\ref{A}. Collecting the above terms we find that $\We$ is given by
\begin{multline}
\We = \frac{\mu}{4}\int_V \int_V \big(\frac{k^2}{\imp^2}\Jmo\cdot \Jmt^* + (\nabla_1\cdot\Jeo)(\nabla_2\cdot\Jet^*)\big)\frac{\cos(kR_{12})}{4\pi k^2 R_{12}}\\ 
+\RE\big\{\big(\frac{1}{\imp^2}[k^2\Jmo\cdot\Jmt^* - (\nabla_1\cdot \Jmo)(\nabla_2\cdot\Jmt^*)] + [k^2\Jeo\cdot\Jet^*-(\nabla_1\cdot\Jeo)(\nabla_2\cdot\Jet^*)]\big)g(\vr_1,\vr_2)\big\} \\
+\frac{2k}{\imp} \IM \big\{(\Jmt^*\times \Jeo)\cdot \nabla_2 g(\vr_1,\vr_2)\big\}\diff V_1\diff V_2.
\end{multline}

Upon inserting the explicit expression for $g$ from~\eqref{ggg} in Appendix~\ref{A}, we arrive to the source representation of the far-field subtracted stored electrical energy: 
\begin{equation}~\label{theWe}
\We= \Wet^{(0)} + \Wem^{(1)} + \Wem^{(2)}+\Wem^{(3)} + \Wem^{\text{rest}} .
\end{equation}
Let the core part of these energies be denoted with $\Wec$, \ie $\Wec=\Wet^{(0)} + \Wem^{(1)} + \Wem^{(2)}+\Wem^{(3)}$.
Of the above terms, only $\Wem^\text{rest}$ is potentially coordinate dependent. The remaining terms are coordinate independent. The rest-term is similar to the term given in~\cite{Yaghjian+Best2005} for {\em electric sources only}, the expression below includes both electric and magnetic sources
\begin{align}
\Wet^{(0)} = \frac{\mu}{4k}\IM \big\{\dotp{\vJe}{\Le\vJe}+\frac{1}{\eta^2}\dotp{\vJm}{\Lm\vJm} \big\}, &&
\Wem^{(1)} = \frac{-\mu}{4k\eta}\IM\dotp{\vJe}{\KK_2\vJm},\\
\Wem^{(2)}=\frac{\mu}{4 k}\IM \big\{\dotp{\vJe}{\Lem\vJe} + \frac{1}{\eta^2}\dotp{\vJm}{\Lem\vJm}\big\}, && 
\Wem^{(3)} = \frac{-\mu}{4\eta k}\RE \dotp{\vJe}{\KK_1\vJm},\\
\Wem^{\text{rest}}= \frac{\mu}{4}\big\{\dotp{\vJe}{\KK_3\vJe} + \frac{1}{\eta^2}\dotp{\vJm}{\KK_3\vJm}+ \dotp{\vJe}{\KK_4\vJm}\big\}, && \label{rest}
\end{align}
where $\dotp{\va}{\vb}=\int_V \va^*\cdot\vb\diff V$ is the dual pairing. 
The sum of the above far-field subtracted stored electrical energy constituents are identical with the definition of the far-field subtracted stored electrical energy~\eqref{WEM}. The constituents of this sum are not unique. We could for example let $\Wem^{(3)}$ be embedded as a perturbation term into the $\Wem^{(1)}$ term or alternatively into the $\Wem^{\text{rest}}$ term. However, there are some points that become easy in using the representation~\eqref{theWe}. Note first that for the electric only terms only $\Wet^{(0)}$, $\Wem^{(2)}$ and a part of $\Wem^{\text{rest}}$ contributes. We will also see that all the $\Wem^{(n)}$ terms also appear in $\Wm$, with only a change of sign associated with $\Wem^{(3)}$. We thus prefer this separation of the sum as given in~\eqref{theWe}. Similarly the radiated power~\eqref{rad} becomes easier to describe if we define the operators $\KK_1$ and $\KK_2$ separately. 

The far-field subtracted stored electrical energy associated operators are given by:
\begin{align}\label{Le}
\dotp{\vJ}{\Le\vJ} &= \frac{-1}{\ju k}\int_V\int_V\big( \nabla_1\cdot \vJ(\vr_1)\big)\big(\nabla_2\cdot \vJ^*(\vr_2)\big)G(\vr_1-\vr_2)\diff V_1\diff V_2,\\\label{Lm}
\dotp{\vJ}{\Lm\vJ} &= \ju k \int_V\int_V  \vJ(\vr_1)\cdot \vJ^*(\vr_2)G(\vr_1-\vr_2)\diff V_1\diff V_2,\\
\dotp{\vJ}{\Lem\vJ}&= -\ju \int_V \int_V  \big(k^2 \vJ_{1}\cdot\vJ_{2}^* - (\nabla_1\cdot\vJ_{1})(\nabla_2\cdot\vJ_{2}^*)\big) \frac{\sin(kR_{12})}{8\pi} \diff V_1\diff V_2 ,\\
\dotp{\vJe}{\KK_1\vJm}&= \frac{k^2}{4\pi}\int_V\int_V \vJe(\vr_1)^*\cdot \hR_{12}\times \vJm(\vr_2)\sbj{1}(kR_{12})\diff V_1\diff V_2,\label{K1}\\
\dotp{\vJe}{\KK_2\vJm}&= \frac{k^2}{4\pi}\int_V\int_V \vJe(\vr_1)^*\cdot \hR_{12}\times \vJm(\vr_2)\cos(k R_{12})\diff V_1\diff V_2 ,\\
\dotp{\vJ}{\KK_3\vJ}&= -\int_V \int_V  \IM\big\{k^2 \vJ_1\cdot\vJ_2^*  -(\nabla_1\cdot \vJ_1)(\nabla_2\cdot\vJ_2^*)\big\} \frac{(r_1^2-r_2^2)}{8\pi R_{12}}\sbj{1}(kR_{12})   \diff V_1\diff V_2, \\
\dotp{\vJe}{\KK_4\vJm}&= \frac{k}{\imp} \int_V \int_V \RE\{\Jmt^*\times \Jeo\} \cdot 
\Big( \frac{ \vr_2+\vr_1}{4\pi R_{12}}\sbj{1}(kR_{12}) + k\hR_{12}\frac{r_1^2-r_2^2}{4\pi R_{12}}\sbj{2}(kR_{12}) \Big)\diff V_1\diff V_2. \label{K4}
\end{align}
Recall that we use the notation of indices $1,2$ on the current densities to indicate a spatial dependence with respect to $\vr_1$, $\vr_2$, {\it viz}: $\vJ_2=\vJ(\vr_2)$, for some position vector $\vr_2$. 
Here $\hR_{12}=\vR_{12}/R_{12}$.

Note that the operators $\Le$ and $\Lm$ are the EIFE operators, which simplifies the implementation of them. 
An alternative way to think about $\KK_1$ is as an element of the magnetic field integral equation kernel, since its integrand is proportional to $-2\ju \Jmt\cdot\nabla_1\IM G\times \Jeo^*$, a similar approach also works to represent $\KK_2$. Another feature of the above expressions are that each of the operators $\Lem$, $\KK_1,\KK_2,\KK_3$, and $\KK_4$ are non-singular which simplifies their numerical implementation. 

We now return to the questions of what type of domain $V$ that can be allowed. 
First, observe that both volume and surface domains are acceptable by utilizing  $J\diff V$ or $\vJ_s\diff S$ in~\eqref{Le}--\eqref{K4}. 
The choice of domain is hence coupled with the considered space of current densities. An implicit definition of the allowed functions $\vJe,\vJm$ and domain $V$ that are acceptable is that \eqref{Le}--\eqref{K4} should be well defined and the space of functions should be rich enough that it is complete. The latter requirement is used implicitly in the optimization process in Sec.~\ref{sec:opt} below to work. For similar considerations for charge densities see \eg~\cite{Landkof1972}.

\subsection{Source representation for the far-field subtracted stored magnetic energy and the definition of stored energies}

To derive the corresponding magnetic far-field subtracted energy we can proceed on a few different equivalent approaches: the first is to repeat the above process, but now for the $\vH$-field. A second alternative is to use the duality invariance of Maxwell's equations with respect to a change of current densities and a third alternative is to use Poynting's theorem~\eqref{PT} to determine the magnetic far-field subtracted stored energy. 

The duality transform of the sources $(\eta\vJe,\vJm)\rightarrow (-\vJm,\eta\vJe)$ gives the far-field subtracted stored magnetic energy as:  
\begin{equation}\label{theWm}
\Wm = \Wmt^{(0)} +  \Wem^{(1)} + \Wem^{(2)} - \Wem^{(3)} + \Wem^{\text{rest}} .
\end{equation}
Let here $\Wmc$ be the core part of the far-field subtracted stored magnetic energy \ie $\Wmc=\Wmt^{(0)} +  \Wem^{(1)} + \Wem^{(2)} - \Wem^{(3)}$.
The first term under the duality transform becomes: 
\begin{equation}
\Wmt^{(0)} = \frac{\mu}{4k}\IM \big\{\dotp{\vJe}{\Lm\vJe}+\frac{1}{\eta^2}\dotp{\vJm}{\Le\vJm} \big\} .
\end{equation}
That $\Wem^{(2)}$ remains unchanged is clear. 
To see that $\Wem^{(3)}$ changes sign, insert the dual sources $(-\vJm/\eta,\vJe)$ into $\Wem^{(3)}$ and note that  
\begin{equation}
[\Wem^{(3)}]_{\text{dual}} = -(\frac{-\mu}{4\eta k})\RE \dotp{\vJm}{\KK_1\vJe} =  \frac{\mu}{4 k \eta}\RE \dotp{\vJe}{\KK_1\vJm} = - \Wem^{(3)}.
\end{equation}
We thus return to the original $\Wem^{(3)}$ expression, but with a change of sign. The terms $\Wem^{(1)}$ remain unchanged under the duality procedure since $-\IM\dotp{\vJm}{\KK_1\vJe} = \IM\dotp{\vJe}{\KK_1\vJm}$. Similarly, $\Wem^\text{rest}$ is unchanged. 

The approach using Poynting's theorem utilize the complex power $\cP$ as defined in~\eqref{cP}. We find that 
\begin{equation}\label{diff}
\Wm-\We = \frac{1}{4}\int_{\RR_r^3} \mu|\vH|^2 - \eps|\vE|^2\diff V = \RE\big\{ \frac{1}{2\ju\omega}\cP\big\}.
\end{equation}
The explicit current-density expression representation for $\cP$, which is given in~\eqref{Ief} in the next section simplify~\eqref{diff} to once again yield the result in~\eqref{theWm}.

Given $\We$ and $\Wm$ we are now in the position to define the stored energies $\Wet$ and $\Wmt$ as the non-negative and coordinate independent core part of $\We$ and $\Wm$:
\begin{equation} \label{Wemt}
\Wet = \max(\Wec,0), \ \ \text{and} \ \ \
\Wmt = \max(\Wmc,0).
\end{equation}
Note that there is also an alternative choice of energy definition: to choose the non-negative and coordinate independent part of $\We$ (and $\Wm$).
We will see the difference between these two concepts becomes important later on.  The reason that we do not define $\Wet$ as the coordinate independent part of $\We$ (and similarly for $\Wmt$) is that $\Wem^{\text{rest}}$ in general contains both a coordinate independent and a coordinate dependent part.  
Separation of them is non-trivial to implement in convex optimization approaches to determine lower bounds on the Q factor.  
Thus, one attractive property of $\Wet$ and $\Wmt$ is that they are straight forward to add to a convex optimization process, as illustrated in Sec.~\ref{sec:opt}.  The definitions in~\eqref{Wemt} still suffers from shortcomings, mainly the requirement of a maximum to ensure that the stored energies are positive. However as the structure grows larger both $\Wet$ and $\Wmt$ can become zero.  
This phenomena is the remainder of that we have subtracted the far field over the entire space, \cf~\eqref{WEM}, including in regions where the field is small or zero. 
For structures of $ka\sim 0.8$ we find that the positive part of $\Wec$ and $\Wmc$ are excellent in estimating the stored energies as compared to the coordinate independent part of $\We$, $\Wm$. 
Below we use $Q$ to indicate that we have used $\Wet$ and $\Wmt$, and $\QF$ when we use $\We$ and $\Wm$.


\section{Radiation intensity and radiated power in terms of sources}\label{sec:Power}

To derive the explicit source representation of the radiated power $\rP$, we utilize its relation to the complex power $\cP$ see~\eqref{rad} and~\eqref{cP}. By a substitution of the fields as given by \eqref{pp} into \eqref{cP} we find the complex power in terms of sources and potentials as:
\begin{multline}\label{4.1}
\cP= \frac{-1}{2}\int_{\RR^3} \vE\cdot \vJe^* + \vH^*\cdot \vJm\diff V  \\ 
=\frac{-1}{2}\int_{\RR^3} \ju\omega (\phi\roe^* -\vA\cdot \vJe^*-\psi^* \rom +\vN^*\cdot \vJm)  - \frac{1}{\eps}\vJe^*\cdot \Rot\vN + \frac{1}{\mu}\vJm\cdot\Rot\vA^* \diff V.
\end{multline}
Here we have used the identity  $\Div(\phi\vJe)=\phi\Div\vJe+\vJe\cdot\nabla \phi$, and the continuity equations for $\vJe$ and $\vJm$. The divergence term $\Div(\phi\vJe)$ vanish due to the bounded support of the source term $\vJe$. 
Replacing all potentials in~\eqref{4.1} with their corresponding Green's function representation~\eqref{AG}--\eqref{FG} yields:
\begin{multline}\label{Ie}
\cP 
= \frac{1}{2}\int_V \int_V \frac{\ju \imp}{k}\Big(k^2 G_{12}\Jeo \cdot \Jet^* -G_{12}(\nabla_1\cdot\Jeo)\nabla_2\cdot \Jet^* +\frac{1}{\imp^2 }G_{12}^*(\nabla_1\cdot\Jmo^*) \nabla_2\cdot \Jmt -\frac{k^2}{\imp^2} G_{12}^*\Jmo^*\cdot \Jmt \Big)  \\ 
-(\Jmo\times\nabla_2 G_{12} ) \cdot \Jet^* + (\Jeo^*\times\nabla_2 G_{12}^* )\cdot \Jmt \diff V_1\diff V_2.
\end{multline}
The last two terms are simplified by using permutation rules for expressions of the form $\va\cdot(\vb\times\vc)$ together with a change of variables $\vr_1\leftrightarrow \vr_2$ in the integrand to become:
\begin{equation}
-\big(\nabla_1 (G_{12}-G_{12}^*)\big)\cdot \Jeo^*\times \Jmt 
= 2\ju \frac{k^2}{4\pi}\Jeo^*\cdot \hR_{12}\times \Jmt\sbj{1}(kR_{12}),
\end{equation}
where $\sbj{n}$ is the spherical Bessel function of order $n$. 
Closer inspection shows that~\eqref{Ie} have large similarities with the electric field integral equation (EFIE). Recall that the EFIE can be expressed in terms of an operator $\LL=\Lm-\Le$ \cite{Jin2011}, see~\eqref{Le} and \eqref{Lm}.
Thus similarly to the EFIE we propose to use these operators to express the integral $\cP$ as
\begin{equation}\label{Ief}
\cP = \frac{\eta}{2}\dotp{\vJe}{\LL\vJe}  + \frac{1}{2\eta}\dotp{\LL\vJm}{\vJm} + \ju \dotp{\vJe}{\KK_1\vJm}.
\end{equation}
The operator $\KK_1$ ia defined in~\eqref{K1}.

From the expression \eqref{Ief} of the complex power $\cP$, we obtain the radiated power by $\rP=\RE\cP$, see~\eqref{rad}. We have thus arrived to the source representation of the radiated power: 
\begin{multline}\label{Prad}
\rP  
= \imp\int_V \int_V \Big(k^2 \Jeo \cdot \Jet^*
-(\nabla_1\cdot \Jeo)\nabla_2\cdot \Jet^*  \\
+\frac{1}{\imp^2}\big(k^2 \Jmo^*\cdot \Jmt-(\nabla_1\cdot \Jmo^*) \nabla_2\cdot \Jmt \big)\Big)\frac{\sin(kR_{12})}{8\pi k R_{12}}\diff V_1\diff V_2 
\\ + \frac{k^2}{4\pi}\int_V \int_V   \sbj{1}(kR_{12})\hat{\vR}_{12} \cdot \IM (\Jeo^*\times\Jmt) \diff V_1\diff V_2
\end{multline}
in the current density.
Note that $\RE\dotp{\vJ}{\LL\vJ} = \RE\dotp{\LL\vJ}{\vJ}$. Utilizing the EFIE operators~\eqref{Le} and \eqref{Lm} we find that the radiated power becomes:
\begin{equation}\label{theP}
\rP = \frac{\eta}{2}\RE \dotp{\vJe}{\LL\vJe}  + \frac{1}{2\eta}\RE\dotp{\vJm}{\LL\vJm} - \IM \dotp{\vJe}{\KK_1\vJm}.
\end{equation}
The above expression for $\rP$ is the desired source term formulation for the radiated power. The first terms containing $\vJe$ is identical to the expression given in \cite{Vandenbosch2010,Gustafsson+Jonsson2015b}. The second two terms containing $\vJm$ are the magnetic dual version of the electric sources. The last integral is a cross term between the electric and magnetic current densities. 

The radiation intensity, $P(\hr)$, for electric and magnetic sources in the direction $\hr$ is given from Poynting's vector as
\begin{equation}\label{UPa}
P(\hr)=\lim_{r\rightarrow \infty}\frac{1}{2} r^2\hr\cdot\RE \{\vE\times \vH^*\}=\lim_{r\rightarrow\infty} \frac{\imp r^2}{2}|\vH|^2
= \lim_{r\rightarrow \infty} \frac{\imp r^2}{2}|-\nabla \psi - \ju\omega \vN + \frac{1}{\mu}\Rot\vA|^2.
\end{equation}
Utilizing the far-field expansion of the Green's solution $G(\vr,\vr_1)\rightarrow \frac{1}{4\pi r}\lexp{-\ju k r + \ju k \hr\cdot \vr_1}$ as $r\rightarrow\infty$ together with the transform of the continuity equation $\int_V \lexp{\ju k \hr\cdot \vr_1}\rom(\vr_1)\diff V_1=\sqrt{\eps\mu}\hr \cdot\int_V \lexp{\ju k \hr\cdot \vr_1}\vJm(\vr_1)\diff V_1$ we find that 
\begin{multline}\label{UP}
P(\hr)=\frac{\imp k^2}{32\pi^2}\Big|\int_V \lexp{\ju k\hr\cdot \vr_1}\big(\frac{1}{\eta}\hr\times(\hr\times \vJm(\vr_1)) - \hr\times \vJe(\vr_1)\big)\diff V_1\Big|^2 \\ = \frac{\eta k^2}{32\pi^2}\Big( \Big|\int_V (\he^*\cdot \vJe(\vr_1) + \frac{1}{\eta}\hr\times \he^* \cdot \vJm(\vr_1))\lexp{\ju k \hr\cdot\vr_1} \diff V_1 \Big|^2 + \Big|\int_V (\hh^*\cdot \vJe(\vr_1) + \frac{1}{\eta}\hr\times \hh^*\cdot \vJm(\vr_1) )\lexp{\ju k \hr\cdot\vr_1} \diff V_1\Big|^2 \Big)  \\ =P(\hr,\he) + P(\hr,\hh).
\end{multline}
In the last step we used that the $\hr,\he^*,\hh^*$ is an orthogonal triplet with $\hr\times \he^*=\hh^*$. We recognize the radiation intensity $P(\hr,\he)$ for a given polarization $\he$. For electric only sources compare with \eg~\cite{Gustafsson+etal2012a}.

We can hence write the partial gain of an antenna in terms of the sources as:
\begin{equation}
G(\hr,\he) = 4\pi\frac{P(\hr,\he)}{\rP} = \frac{\eta k^2}{4\pi} \frac{\big|\int_V (\he^*\cdot \vJe(\vr_1) +\frac{1}{\eta}\hr\times \he^*\cdot \vJm(\vr_1))\lexp{\ju k\hr\cdot \vr_1}\diff V_1\big|^2}{\eta\RE \dotp{\vJe}{\LL\vJe}  + \frac{1}{\eta}\RE\dotp{\vJm}{\LL\vJm} - 2\IM \dotp{\vJe}{\KK_1\vJm}}.
\end{equation}

\section{Asymptotic behavior when the support is electrically small}\label{sec:small}

We have above assumed that the support of the electric and magnetic sources is a bounded region. To derive the expressions for the stored energies there has been no need to make assumptions on having an electrically small region. In this section, we briefly discuss the limit of electrically small regions, \ie $ka\rightarrow 0$, where $a$ is the radius of an enclosing sphere, see Fig.~\ref{support}.  A thorough discussion the electrically small case can be found in~\cite{Jonsson+Gustafsson2015}. We assume, only in this section, that both the electric and magnetic current densities have the asymptotic expansion $\vJ=\vJ^{(0)}+k\vJ^{(1)}$ as $k\to 0$, where $\Div\vJ^{(0)}=0$. Then we note that the terms denoted $\Wem^{(n)}$ in the far-field subtracted stored energy terms $\We$ and $\Wm$ have an asymptotic behavior such that $\Wem^{(n)}$ is bounded by a constant times $(ka)^n$, for $n=1,2,3$ as $ka\rightarrow 0$. Similarly we have that both $\Wet^{(0)}$ and $\Wmt^{(0)}$ are bounded by a constant as $ka\rightarrow 0$. The coordinate dependent term $\Wem^{\text{rest}}$ is bounded by  a constant times $(ka)^3$ as $ka\rightarrow 0$. Thus to express the antenna Q for an electrical dipole radiator with polarization in the $\he$-direction in terms of polarizabilities, we use that $\We = W_{\mrm{e},0}+\Oh(ka)$ as $ka\to0$ and similarly for $\Wm$ to find~\cite{Yaghjian+etal2013,Jonsson+Gustafsson2015a} 
\begin{align}\label{Qza}
Q_{\mrm{e},0} & = \min_{\vJe^{(1)}} \frac{2\omega W_{\mrm{e,0}}}{\rP}= \frac{6\pi}{k^3 \he\cdot\gmate\cdot \he}, \ \ \
\ Q_{\mrm{m},0}= \min_{\vJm^{(0)}} \frac{2\omega W_{\mrm{e,0}}}{\rP}= \frac{6\pi}{k^3 \he\cdot\gmatm\cdot \he}, \\
\ Q_{\mrm{em},0} & = \min_{\vJm^{(0)},\vJe^{(1)}} \frac{2\omega W_{\mrm{e,0}}}{\rP}= \frac{6\pi}{k^3 \he\cdot(\gmate+\gmatm)\cdot \he}.\label{Qzb}
\end{align}
Equivalently we could have obtained $Q_{\mrm{m},0}$ by optimizing $2\omega W_{\mrm{m},0}/\rP$ over $\vJe^{(0)}$ since they are dual symmetric.
The minimizations comes with the constraints that $\hn \cdot \vJe^{(1)}=0$ and $\hn\cdot \vJm^{(0)}=0$ over the surface and that $\Div\vJm^{(0)}=0$, see~\cite{Jonsson+Gustafsson2015} for a discussion. Here $\hn$ is the normal to the surface at a given point. 
 Here $\gmate$ and $\gmatm$ are the static electric and magnetic polarizabilities respectively see \eg~\cite{Jones1985,Sjoberg2009b} for their properties. It was shown in~\cite{Jonsson+Gustafsson2015} that the above minimization problem are the same as the general minimization problems in the limit $ka\rightarrow 0$ with the assumption that the structure radiates as an electric dipole. 

In contrast to the different growth rates of the terms in the stored energy we have that all terms in $\rP$ as given by~\eqref{theP} are bounded by a constant times $(ka)^4$ as $ka\rightarrow 0$. It is perhaps somewhat surprising that all terms in $\rP$ are of the same order, but that the cross-term in $\We$ and $\Wm$ are of lower order, in view of the relation~\eqref{diff} and $\rP=\RE{\cP}$. The cause to this difference can be found in that the real and imaginary part of $\cP$ have different size as $ka\rightarrow 0$. Similar behavior are common, consider \eg the small $ka$ behavior of the real and imaginary part of $\lexp{-\ju k a}/(ka)$. 
Further discussion of both electrically small supports of antennas and their current structures can be found in~\cite{Jonsson+Gustafsson2013a,Jonsson+Gustafsson2015a,Jonsson+Gustafsson2015}. 

\section{Coordinate dependence of the far-field subtracted stored energies}\label{Coord}

The far-field subtracted stored energies defined in~\eqref{theWe} and \eqref{theWm} suffer from a weak coordinate dependence similar to earlier approaches to define stored energies, see~\eg~\cite{Collin+Rothschild1964,Fante1969,Yaghjian+Best2005,Gustafsson+Jonsson2015b}, as a difference to $\Wet$ and $\Wmt$ that are coordinate independent. To explicitly determine the dependence of $\We$ we shift the coordinate origin a distance $\vd$ and find that the change of the far-field subtracted stored electrical energy $\delta \We=(\We)_d-(\We)_0$ is
\begin{equation} 
\delta \We = \delta\Wem^{\text{rest}} = \frac{\mu}{4}\big(\dotp{\vJe}{\delta\KK_3\vJe}+\frac{1}{\imp^2}\dotp{\vJm}{\delta\KK_3\vJm}+\dotp{\vJe}{\delta\KK_4\vJm}\big).
\end{equation}
The change in $\KK_3$ is given by
\begin{equation}
\dotp{\vJ}{\delta\KK_3\vJ}= -2\vd\cdot \IM \int_V\int_V \big(k^2\vJ_1\cdot\vJ_2^*-(\nabla_1\cdot\vJ_1)(\nabla_2\cdot\vJ_2^*)\big)\frac{\hR_{12}}{8\pi}\sbj{1}(kR_{12})\diff V_1\diff V_2\diff \Omega.
\end{equation}
Utilizing the integral identity~\eqref{int1} we find that 
\begin{multline}
\dotp{\vJ}{\delta\KK_3\vJ}= 
\frac{-1}{16\pi^2} \RE \int_V\int_V \big(k^2\vJ_1\cdot\vJ_2^*-(\nabla_1\cdot\vJ_1)(\nabla_2\cdot\vJ_2^*)\big)\int_{\Omega} \hr\cdot\vd\lexp{\ju k\hr\cdot (\vr_1-\vr_2)}\diff \Omega\diff V_1\diff V_2  \\
=\frac{-1}{16\pi^2} \RE \int_\Omega \vd\cdot \hr \Big(k^2\big|\int_V \lexp{\ju k\hr\cdot\vr_1}\vJ(\vr_1) \diff V_1\big|^2 -\big| \int_V \lexp{\ju k \hr\cdot \vr_1 }(\nabla_1\cdot\vJ_1)\diff V_1\big|^2 \Big) \diff \Omega.
\end{multline}
The definition of the far-fields amplitudes~\eqref{Ao}-\eqref{psio} gives us that 
\begin{equation}
\frac{\mu}{4}\big(\dotp{\vJe}{\delta\KK_3\vJe}+\dotp{\vJm}{\delta\KK_3\vJm}\big)=
\frac{-\eps}{4}\int_{\Omega}\vd\cdot\hr (\omega^2|\Ao|^2-k^2|\phio|^2+\frac{k^2}{\eps^2}|\Fo|^2-k^2\eta^2|\psio|^2)\diff \Omega.
\end{equation}
Similarly we have from $\delta\KK_4$ that 
\begin{equation}
\dotp{\vJe}{\delta\KK_4\vJm} = 
\frac{k^2}{4\pi\imp} \int_V \int_V \RE \{\Jmt^*\times \Jeo\} \cdot 
\big( \frac{ 2\vd}{ k R_{12}}\sbj{1}(kR_{12}) + \hR_{12}(2\vd\cdot \hR_{12})\sbj{2}(kR_{12}) \big)\diff V_1\diff V_2. 
\end{equation}
Upon utilizing the integral identity~\eqref{int2} we find that 
\begin{equation}
\dotp{\vJe}{\delta\KK_4\vJm} = \frac{k^2}{\eta\eps\mu} 2\vd\cdot\int_{\Omega} \hr \RE\{\Ao\cdot\hr\times\Fo^*\} \diff\Omega.
\end{equation}
Collecting terms and comparing with~\eqref{Eo} we find that 
\begin{equation}\label{eq:coordshift}
\delta\We = \frac{-\eps}{4}\int_\Omega \vd\cdot\hr|\Eo|^2\diff \Omega.
\end{equation}
Thus the shift of coordinates from $\vr\mapsto\vr+\vd$ gives the same change of the far-field subtracted stored energies as given in~\cite{Gustafsson+Jonsson2015b} derived for electric only currents. The expression also have the same character as the coordinate dependence derived in~\cite{Yaghjian+Best2005}. 

\begin{figure}[t!]
\centering
\includegraphics[width=.9\textwidth]{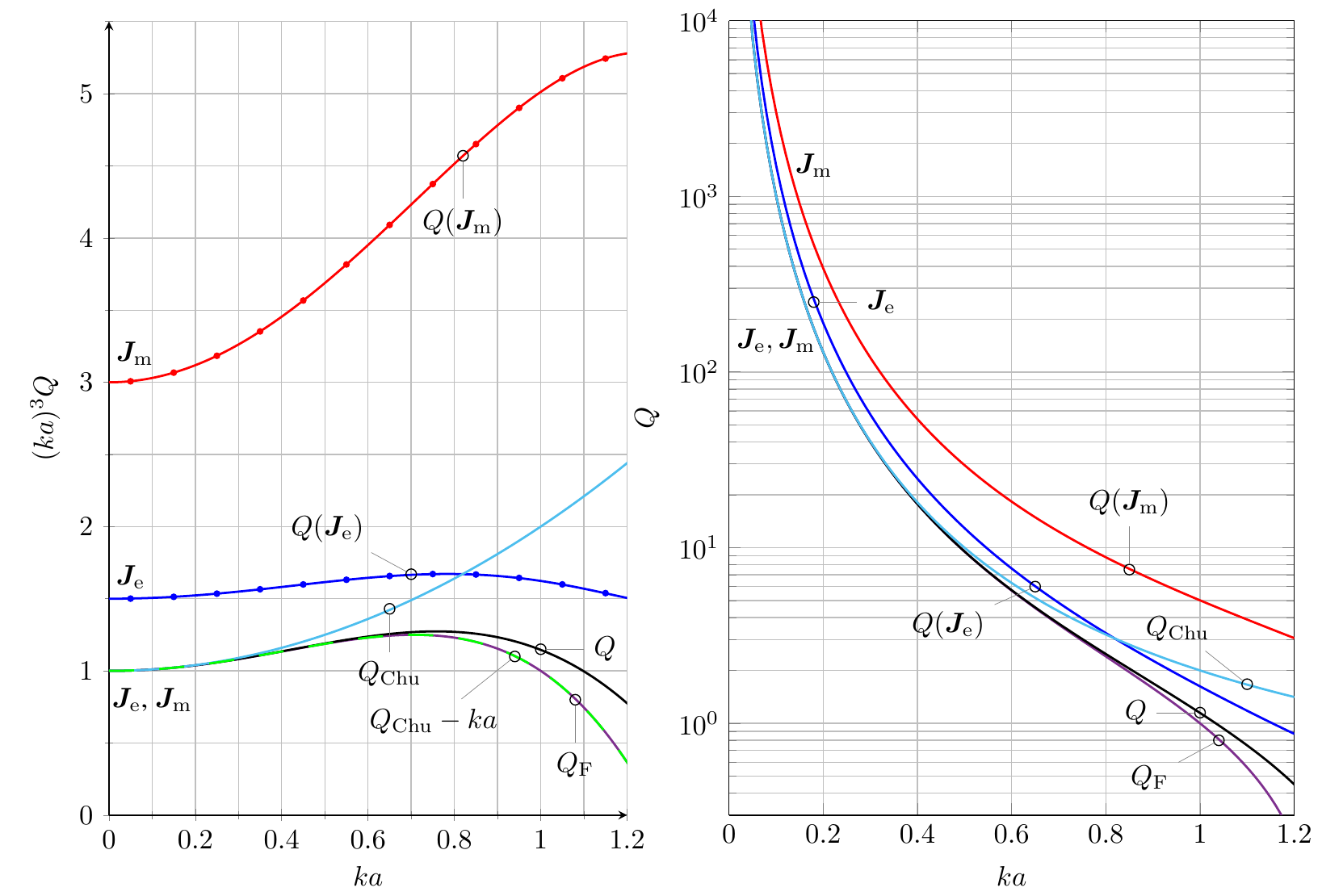}
\caption{The figure depicts lower bound on antenna Q for a sphere with the condition that the field radiates as an electric dipole. The three cases marked with $Q(\vJm)$, $Q(\vJe)$ and $Q$ correspond to only $\vJm$-sources, only $\vJe$-source, and both electric and magnetic sources on the surface respectively. In the figure, we have also included Chu's result marked with $Q_\mrm{Chu}$, see~\eqref{eq:Chubound} and $Q_\mrm{Chu}$ subtracted with $ka$ to account for the removal of the far-field within the sphere given in~\eqref{WEM}. The curve marked with $\QF$ corresponds to the far-field subtracted stored energies~\eqref{WEM} and it cannot be distinguished from $Q_{\text{Chu}}-ka$. The energies $\Wet$ and $\Wmt$, see~\eqref{Wemt}, give rise to the curve marked with $Q$. Observe that $\Qc$ and $\QF$ deviate above $ka\sim 0.8$.}\label{fig:chu}
\end{figure}

We use the stored energies $\Wet$ \eqref{theWe} and  $\Wmt$ \eqref{theWm} inserted in the definition of the Q factor \eqref{eq:Qdef} to determine a lower bound for the sphere, that radiates as an electric dipole. This is compared with the lower bound on the Q factor by Chu~\cite{Chu1948}:
\begin{equation}\label{eq:Chubound}
Q_\text{Chu} = \frac{1}{(ka)^3} + \frac{1}{ka}.
\end{equation}
The electric and magnetic surface current densities $\vJ_{\mrm{es}}=J\sin\theta\hth$, and $\vJ_{\mrm{ms}}=K\sin\theta\hphi$ correspond to electric dipole radiation associated with a dipole in the $\hz$-direction. Dipole-radiation is the lowest radiating mode(s) on a sphere.  Here $\varphi$ is the azimuth angle and $\theta$ is the angle from the $z$-axis, and $\hth$ and $\hphi$ are the standard spherical unit-vectors in the directions of $\theta$ and $\varphi$ respectively.  To find the far-field subtracted stored energies we calculate analytically each of the integrals given in~\eqref{Le}--\eqref{K4} for these current densities. We also calculate $\rP$ explicitly.  For pure magnetic current densities that radiate as an electrical dipole we find the lowest antenna Q depicted in Fig.~\ref{fig:chu} and marked with $\vJm$.  
Similarly for purely electric current densities that radiate as an electrical dipole, we find the lower bound as marked with $\vJe$ in Fig.~\ref{fig:chu}.  To study the case with both electric and magnetic current densities, we need to determine the ratio between the electric and magnetic current densities.  
It is here optimized for each value of $ka$ to find the lowest antenna Q for currents that radiate as an electrical dipole. The result is given as the graph marked with $\vJe$ and $\vJm$. 
To determine these antenna Q's we need to optimize the relation between $J$ and $K$ to find the lowest possible $Q$.  We calculate the antenna Q with and without the $\Wem^\text{rest}$-term as marked with $\QF$ and $Q$ in Fig.~\ref{fig:chu}, respectively. With the rest term, the field inside the sphere becomes zero as for the case by Chu~\cite{Chu1948}. Without the rest term, there is a small non-zero field inside the sphere for the optimal $Q$-value.  We note that the rest-term contributes to lower the stored energies for the combined current density case.  The rest-term is thus essential to obtain the Chu-limit, this difference appears first at $ka\sim 0.8$, where $Q$ is rather small.  
Observe also that $Q$ for electric, magnetic, and combined sources, scales differently as a function of $ka$ in Fig.~\ref{fig:chu}. It is clear that $Q(\vJe)k^3a^3$ is much flatter than the two other cases. A similar type of scaling is observed also in the numerical examples in Sec.~\ref{sec:Ill}. This is also interesting when we compare with the leading order bounds of $Q$ that are expressed in the polarizabilities corresponding to $\{1, 1.5, 3\}(ka)^{-3}$ for the sphere, see~\eqref{Qza}-\eqref{Qzb} and \cite{Jonsson+Gustafsson2015}.

\section{Matrix formulation of energies and radiated power}~\label{sec:Matrix}
To simplify the numerical implementation of the quantities $\Wet$, $\Wmt$ and $\rP$, we consider them as quadratic forms of $6\times 6$-matrices of operators. Introduce the notation $\JS=(\vJe,\eta^{-1}\vJm)$ and the corresponding dual pairing $\dotp{\JS}{\mathcal{Z}\JS}=\int_V \JS^\mrm{H}(\vr_1)(\mathcal{Z}\JS)(\vr_1)\diff V_1$, where some $6\times 6$ matrix $\mathcal{Z}$ of operators act upon function elements $\JS$. Here $\cdot^\mrm{H}$ denote the transpose and complex conjugate. Expansion of the real and imaginary parts in the cross terms yield the off-diagonal terms. 
We note that for a coordinate independent term, here represented by an operator $\mathcal{B}$ with an even kernel, $b(\vr)=b(-\vr)$, 
 we have 
\begin{multline}
2 \ju \IM\dotp{\vJ}{\mathcal{B}\vJ} = 
\int_V \int_V \vJ_1\cdot\vJ_2^* (b(\vr_1-\vr_2)-b(\vr_2-\vr_1)^*)\diff V_1\diff V_2  \\ =   
\int_V \int_V \vJ_1\cdot\vJ_2^* 2\ju \IM\{b(\vr_1-\vr_2)\}\diff V_1\diff V_2  = 
2\ju \dotp{\vJ}{\IM\{\mathcal{B}\}\vJ}.
\end{multline}
For odd kernels $b(\rv)=-b(-\rv)$, we get $2\RE\{\mathcal{B}\}$ and find
\begin{equation}
\Wet = \max(\dotp{\JS}{\mathcal{X}_\mrm{e}\JS},0), \ \ 
\mathcal{X}_\mrm{e} 
= \frac{\eta}{4\omega}
\begin{pmatrix} 
 \IM\{\Le+\Lem\} & -\frac{1}{2\ju}\KK_2 - \frac{1}{2}\KK_1\\ \frac{1}{2\ju}\KK_2 - \frac{1}{2}\KK_1 & \IM\{\Lm+\Lem\}
\end{pmatrix}
\label{eq:We}
\end{equation}
and
\begin{equation}
\Wmt = \max(\dotp{\JS}{\mathcal{X}_\mrm{m}\JS},0), \ \ \mathcal{X}_\mrm{m}
=
\frac{\eta}{4\omega}
\begin{pmatrix} 
\IM\{\Lm+\Lem\} & -\frac{1}{2\ju}\KK_2 + \frac{1}{2}\KK_1 \\ \frac{1}{2\ju}\KK_2 + \frac{1}{2}\KK_1 & \IM\{\Le+\Lem\}
\end{pmatrix}
\label{eq:Wm}
\end{equation}
together with
\begin{equation}
\rP=\dotp{\JS}{\mathcal{R}\JS}, \ \mathcal{R}=
 \frac{\eta}{2}\begin{pmatrix} 
 \RE\{\Lm-\Le\} & \ju \KK_1\\
 -\ju \KK_1 & \RE\{\Lm-\Le\}
\end{pmatrix}.
\label{eq:Prad}
\end{equation}
The duality transformation $(\eta\vJe,\vJm)\mapsto(-\vJm',\vJe'\eta)$ gives that $\Wet\leftrightarrow \Wmt$, and $\rP\leftrightarrow \rP$, as is expected from the invariance of Maxwell's equations under duality transforms. 

We note that the leading order stored energies consist of the EFIE and MFIE-operators, and hence comes as a small extra cost given the MoM matrices. To calculate all parts of the stored energies as well as the radiated power one needs to calculate three additional matrices corresponding to the operators $\Lem$, $\KK_1$, and $\KK_2$, these extra matrices are non-singular, and easy to implement.

\section{Convex optimization problems}\label{sec:opt}

Given a particular volume or shape, we formulate the optimization problem for $\Qc$ and $G/\Qc$ in terms of the current densities. Maximization of $G/\Qc$ is  
\begin{equation}
  \frac{G}{\Qc} = \sup_{\vJe,\vJm} \frac{2\pi P(\hr,\he)}{\omega \max(\Wet,\Wmt)},
\label{eq:QandGQ}
\end{equation}
where the current densities that the optimization is over are the controlled currents, \ie the sources that can be controlled. It is shown through a series of examples in~\cite{Gustafsson+Nordebo2013,Gustafsson+Jonsson2015a,Gustafsson+etal2012a,Cismasu+Gustafsson2014a} that these controlled currents can be used to model the performance of antennas. Note that the function space for the current densities is the space for which the optimized quantity is well defined. 

Consider a region $V$ with the electric $\Jve=\Jve(\rv)$ and magnetic $\Jve=\Jve(\rv)$ current densities. The current densities are expanded in local basis functions $\psiv_n$ as
\begin{equation}\label{eq:basis}
	\Jve(\rv)\approx \sum_{n=1}^N I_{\mrm{e},n}\psiv_n(\rv)
  \qtext{and }
  \Jvm(\rv)\approx \frac{1}{\eta}\sum_{n=1}^N I_{\mrm{m},n}\psiv_n(\rv).
\end{equation}
We introduce the $N\times1$ current matrices $\Ime$ and $\Imm$ with the elements $I_{\mrm{e},n}$ and $I_{\mrm{m},n}$, respectively, to simplify the notation. The electric and magnetic currents are also collected in the $2N\times 1$ current matrix $\Jm$ containing $I_{\mrm{e},n}$ in rows 1 to $N$ and $I_{\mrm{m},n}$ in rows $N+1$ to $2N$. Insertion of the expansions~\eqref{eq:basis} in \eqref{eq:We} and~\eqref{eq:Wm} gives the approximation
\begin{equation}
	\Wet \approx\frac{1}{4\omega}
  \begin{pmatrix}
    \Ime \\ \Imm
  \end{pmatrix}^{\herm}  
  \begin{pmatrix}
    \Xmee+\Xmem & \frac{-1}{2\ju}\Km_2-\frac{1}{2}\Km_1\\
    \frac{1}{2\ju}\Km_2-\frac{1}{2}\Km_1 & \Xmmm+\Xmem
  \end{pmatrix}
  \begin{pmatrix}
    \Ime \\ \Imm
  \end{pmatrix}
  =\frac{1}{4\omega}
  \Jm^{\herm}\Xme\Jm  	
\label{eq:WeI}
\end{equation}
for the stored electric energy and
\begin{equation}
	\Wmt \approx\frac{1}{4\omega}
  \begin{pmatrix}
    \Ime \\ \Imm
  \end{pmatrix}^{\herm}  
  \begin{pmatrix}
    \Xmmm+\Xmem & \frac{-1}{2\ju}\Km_2+\frac{1}{2}\Km_1\\
    \frac{1}{2\ju}\Km_1+\frac{1}{2}\Km_2 & \Xmee+\Xmem
  \end{pmatrix}
  \begin{pmatrix}
    \Ime \\ \Imm
  \end{pmatrix}
  =\frac{1}{4\omega}
  \Jm^{\herm}\Xmm\Jm  	
\label{eq:WmI}
\end{equation}
for the stored magnetic energy, where the reactance matrices $\Xmee$, $\Xmem$, $\Xmme$, $\Xmmm$, $\Km_1$, and $\Km_2$ are introduced and the superscript ${}^\herm$ denotes the Hermitian transpose. 
Note here that we have included $\eta$ into our definition of the matrices $\Xme$, $\Xmm$ etc as compared with the operator expressions given in~\eqref{eq:We} and \eqref{eq:Wm}. The corresponding approximation of the radiated power~\eqref{eq:Prad} and far field are
\begin{equation}
  P_{\mrm{rad}} \approx \frac{1}{2}\Jm^{\herm}\Rm\Jm  	
  \qtext{and }\ \ 
  \evh^{\ast}\cdot\Eo(\rvh)
  \approx \Fm\Jm,
\label{eq:Rm}
\end{equation} 
respectively. Maximization of the quotient between the partial gain and Q-factor~\eqref{eq:QandGQ} is approximated as
\begin{equation}
  \frac{G}{\Qc} \approx \max_{\Jm}\frac{4\pi|\Fm\Jm|^2}{\eta\max\{\Jm^{\herm}\Xme\Jm,\Jm^{\herm}\Xmm\Jm\}}.
\label{eq:GoQm}
\end{equation}
In this paper, we assume that the expansion~\eqref{eq:basis} is sufficiently accurate to replace the approximate equal to in~\eqref{eq:GoQm} with equalities. The matrix formulation~\eqref{eq:GoQm} is formally equivalent with the case having purely electric currents in~\cite{Gustafsson+etal2012a,Gustafsson+Nordebo2013,Gustafsson+etal2016a}. The maximization problem~\eqref{eq:GoQm} is transformed to the convex optimization problem
\begin{equation}\label{eq:GoQ2}
	\begin{aligned}
		& \minimize && \max\{\Jm^{\herm}\Xme\Jm,\Jm^{\herm}\Xmm\Jm\}  \\
		& \subto && \Fm\Jm = 1.
	\end{aligned}
\end{equation}
Here, we have assumed that the matrices $\Xme$ and $\Xmm$ are positive semidefinite, \ie $\Xme, \Xmm\succeq\Om$, that follows from the definitions in~\eqref{Wemt} by replacing negative eigenvalues of $\Xme$ and $\Xmm$ with zeros, see~\cite{Gustafsson+Nordebo2013,Gustafsson+etal2016a}.
The convex optimization can be solved using standard software such as CVX~\cite{Grant+Boyd2011} or Newton's method~\cite{Gustafsson+etal2016a}.

Minimization of the Q-factor~\eqref{eq:QandGQ} is not convex and it has so far not been reformulated as a convex optimization problem. Particular cases such as the limit $ka\to 0$~\cite{Jonsson+Gustafsson2015} can be solved and relaxation can be used for some other cases~\cite{Gustafsson+etal2016a}. Moreover, we are often interested in the minimum Q-factor for a specific radiated field such as linear or circular polarized cases, \cf the classical Chu bound~\eqref{eq:Chubound} for a dipole with the mixed mode case~\cite{Chu1948,McLean1996}. Here, we consider the case with minimization of the Q-factor, where the radiated power in a desired radiated mode is considered~\cite{Gustafsson+Nordebo2013,Gustafsson+etal2016a}. We focus on the dipole modes similar to the classical Chu bound~\cite{Chu1948}. Expand the far-field in spherical modes and let $\Mm_m\Jm$ denote the projection of the current on mode number $m$. The radiated power is expressed as
\begin{equation}
  \Jm^{\herm}\Rm\Jm = \sum_{m=1}^{\infty}|\Mm_m\Jm|^2\geq |\Mm_0\Jm|^2
\label{eq:Proj}
\end{equation}
giving the convex optimization problem
\begin{equation}\label{eq:minQ}
	\begin{aligned}
		& \minimize && \max\{\Jm^{\herm}\Xme\Jm,\Jm^{\herm}\Xmm\Jm\}  \\
		& \subto && \Mm_0\Jm = 1.
	\end{aligned}
\end{equation}
Here $\Mm_0$ is one or a linear combintation of the projections on the dipole-modes in the set $\{\Mm_m\}_{m=1}^\infty$.

\section{Numerical examples}\label{sec:Ill}
Numerical examples illustrating the minimum Q-factor for spherical capped dipole, cylinders, spherical shells are presented in Secs.~\ref{S:ExCappedDipole}, \ref{S:ExCylinder}, and~\ref{S:ExSphShell}, respectively. The convex optimization problem~\eqref{eq:minQ} is solved using CVX~\cite{Grant+Boyd2011}. Alternatively, the Newton's method can be used~\cite{Gustafsson+etal2016a}. The problems are solved for the electrical sizes $ka=\{0,0.1,0.3,0.5\}$ as depicted using the line styles $\{$
\begin{tikzpicture}
\def\dx{0.5}
\def\dd{0.3}
\draw[black,thick]  (-\dx-\dd,0) -- +(\dx,0) node[right] {,};
\filldraw (-\dx/2-\dd,0) circle (1pt);
\draw[blue!50!black,thick]  (0,0) -- +(\dx,0) node[right] {,};
\draw[red!50!black,thick,dashed] (\dx+\dd,0) -- +(\dx,0) node[right] {,};
\draw[green!50!black,thick,densely dotted] (2*\dx+2*\dd,0) -- +(\dx,0);
\end{tikzpicture}
$\}$, respectively, where $a$ denotes the radius of the smallest circumscribing sphere, see Fig.~\ref{support}. The polarizability dyadics are used for the $ka=0$ cases~\cite{Gustafsson+etal2012a,Yaghjian+etal2013,Jonsson+Gustafsson2015}.
The geometries are parametrized with the polar angle $\theta$, see Fig.~\ref{fig:CapSphere}. The geometries are considered together with the surfaces of their convex hull.  

\subsection{Spherical capped dipole}\label{S:ExCappedDipole}
\begin{figure}[tp]%
\centering
\includegraphics[width=0.45\columnwidth]{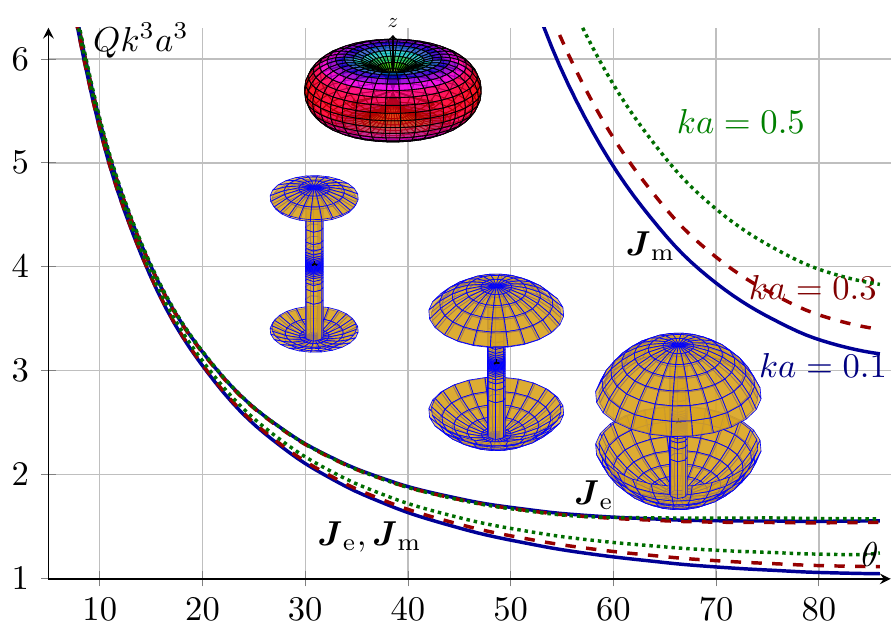}%
\hspace{3mm}
\includegraphics[width=0.45\columnwidth]{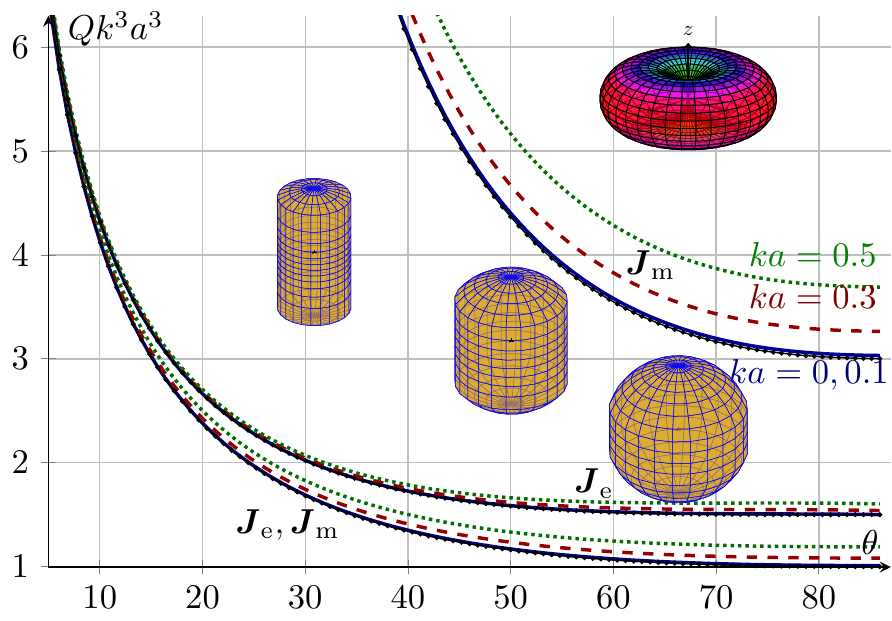}%
\caption{Minimum Q-factors for spherical capped dipole antennas with aperture angle $\theta$ and thickness $a/20$ radiating as an electrical dipole with electric currents, magnetic currents, and combined electric and magnetic currents. The resulting Q-factors are normalized with $k^3a^3$ for the cases $ka=\{0,0.1,0.3,0.5\}$. Antenna structure and its surface convex hull in the left and right figures, respectively.}%
\label{fig:CapSphere}%
\end{figure}

The spherical capped dipole~\cite{Stuart+Pidwerbetsky2006} with different opening angles, electrical sizes, and current excitations are used to illustrate minimum Q-factors. The convex optimization problem~\eqref{eq:minQ} to minimize the Q-factor for an antenna radiation as an electric dipole is solved with electric currents, magnetic currents, and a combination of electric and magnetic currents. The resulting Q-factors normalized with $(ka)^{-3}$ are depicted in Fig.~\ref{fig:CapSphere}. 
 
It is observed that combination of electric and magnetic currents offers the lowest Q-factor. The high Q-factors for the magnetic currents are understood from the explicit solution of the lower bound on the Q-factor in the limit $ka\to 0$ for a linear polarization $\evh$~\cite{Yaghjian+etal2013,Jonsson+Gustafsson2015}
\begin{equation}
  Q_{\mrm{e},0}\geq \frac{6\pi}{k^3\evh\cdot\gmate\cdot\evh},
  \quad 
  Q_{\mrm{m},0}\geq \frac{6\pi}{k^3\evh\cdot\gmatm\cdot\evh},
  \qtext{and }
  Q_{\mrm{em},0}\geq \frac{6\pi}{k^3\evh\cdot(\gmate+\gmatm)\cdot\evh}
  =\frac{1}{Q_{\mrm{e},0}^{-1}+Q_{\mrm{m},0}^{-1}},
\label{eq:Qka20}
\end{equation} 
where $\gmate$ and $\gmatm$ denote the electric and magnetic polarizability dyadics, respectively. The zero subscript on $Q$ indicates that we are only including the leading order term from $\Wet$ and $\Wmt$ and the e-subscript indicates that only electrical currents are used, whereas the m-subscript shows that only magnetic currents are used. The structure radiates as a dipole. The subscript em-has thus both electric and magnetic currents. The bound~\eqref{eq:Qka20} simplifies for a body of revolution antenna with vertical polarization $\evh=\zvh$ to
\begin{equation}
  Q_{\mrm{e},0}\geq \frac{6\pi}{k^3\gamma_{\mrm{zz}}},
  \quad 
  Q_{\mrm{m},0}\geq \frac{12\pi}{k^3\gamma_{\mrm{xx}}},
  \qtext{and }
  Q_{\mrm{em},0}\geq \frac{6\pi}{k^3(\gamma_{\mrm{zz}}+\gamma_{\mrm{xx}}/2)},
\label{eq:Qka20bor}
\end{equation} 
where $\gamma_{\mrm{zz}}$ and $\gamma_{\mrm{xx}}$ denote the high-contrast polarizability~\cite{Gustafsson+etal2007a} in the $\zvh$ and $\xvh$ directions, respectively~\cite{Gustafsson+etal2012a,Jonsson+Gustafsson2015}. 
The high-contrast polarizability is related to the induced dipole moment for the structure immersed in an electrostatic field and hence to the ability of the structure to separate charge~\cite{Gustafsson+etal2007a,Gustafsson+etal2009a}. 
The spherical capped dipole structure has a higher polarizability in the vertical direction than in the horizontal direction due to its ability to store charge on the top and bottom spherical shells. The explicit results~\eqref{eq:Qka20bor} also explains the negligible reduction of the Q-factor when using electric and magnetic currents as $\gamma_{\mrm{xx}}\ll\gamma_{\mrm{zz}}$ for small angles $\theta$. Moreover $\gamma_{\mrm{xx}}=\gamma_{\mrm{zz}}$ for $\theta=90^{\circ}$, where the Q-factors approach the classical bounds by Chu~\cite{Chu1948} and Thal~\cite{Thal2006}.

The decrease of the Q-factor for the case with combined electric and magnetic currents can also be explained by the reduction of the stored energy in the interior part of the structure~\cite{Thal2006}. This is often motivated by equivalent currents that can produce quiescent fields in the interior volume~\cite{vanBladel2007}. The spherical capped dipole is however very thin, here we consider a thickness $a/20$ of the spherical shell and $a/10$ for the diameter of the inner cylinder. It has a negligible interior region so this cannot explain the reduction of the stored energy. Instead the electric and magnetic current reduce the stored energy in a region that is closer to the convex hull of the structure, see~\cite{Jonsson+Gustafsson2015} for an illustration. The corresponding Q-factors for the convex hull of the spherical capped dipole are depicted in Fig.~\ref{fig:CapSphere}. The results are similar to the case for the spherical capped dipole but the Q-factors are slightly lower. The magnetic case has the largest reduction as again is easily explained by the increase of the polarizability in the horizontal direction, \ie for $\gamma_{\mrm{xx}}$.

Fig.~\ref{fig:CapSphere} also shows that the Q-factors depend on the electrical size of $ka$ the object. The $k^3a^3$ scaling dominates the dependence and is very good normalization factor for the most common case for electric dipole type antennas composed of electric currents or similarly electric materials. This is also observed from the forward scattering bounds~\cite{Gustafsson+etal2007a,Gustafsson+etal2009a,Gustafsson+etal2010a} as $D/Q$ is proportional to $k^3a^3$ and $D\approx 1.5$. The dependence is slightly different for the cases with magnetic currents and combined electric and magnetic currents. 

\subsection{Cylindrical regions}\label{S:ExCylinder}
\begin{figure}[tp]%
\centering
\includegraphics[width=0.45\columnwidth]{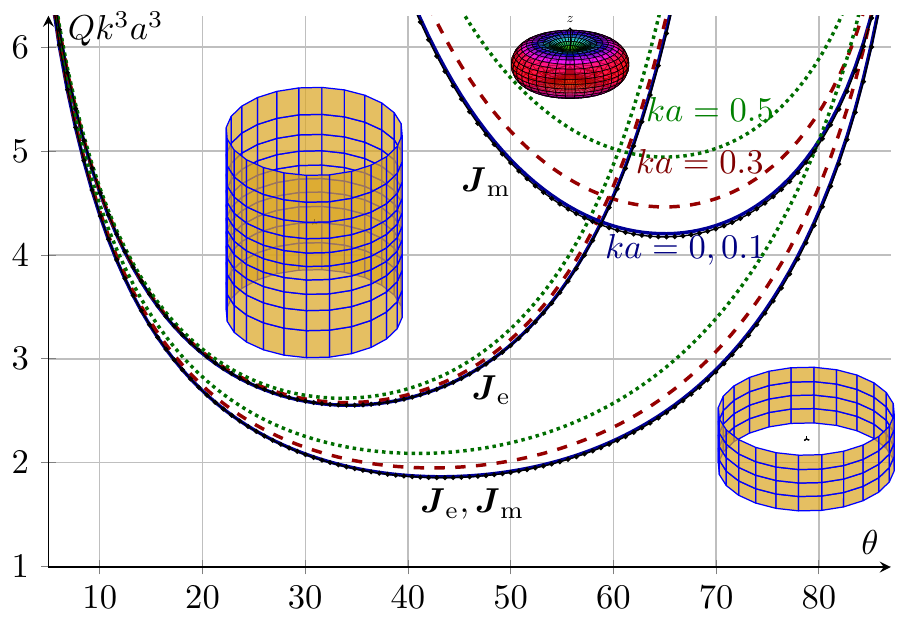}%
\hspace{3mm}
\includegraphics[width=0.45\columnwidth]{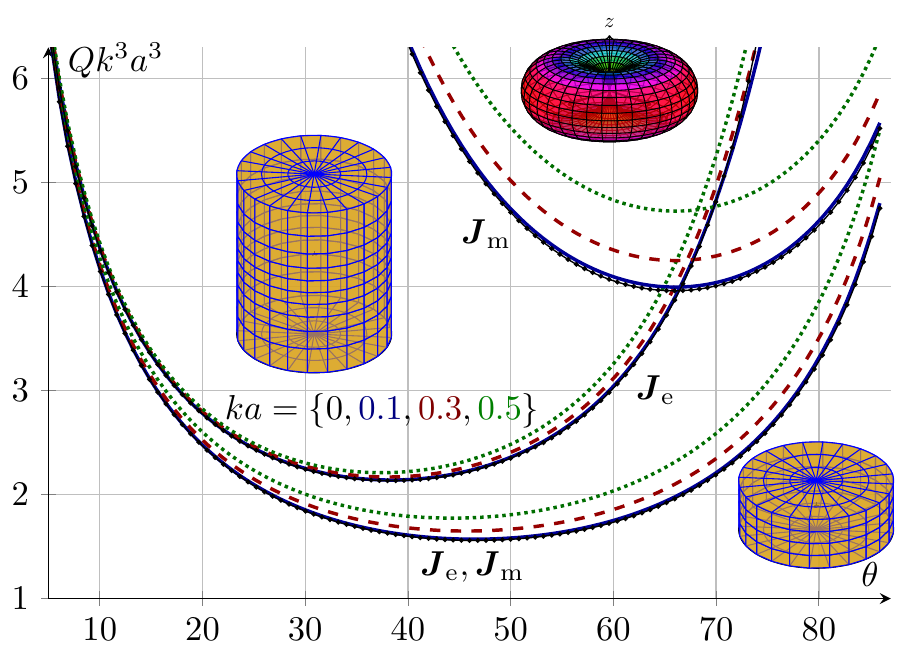}%

\centering
\includegraphics[width=0.45\columnwidth]{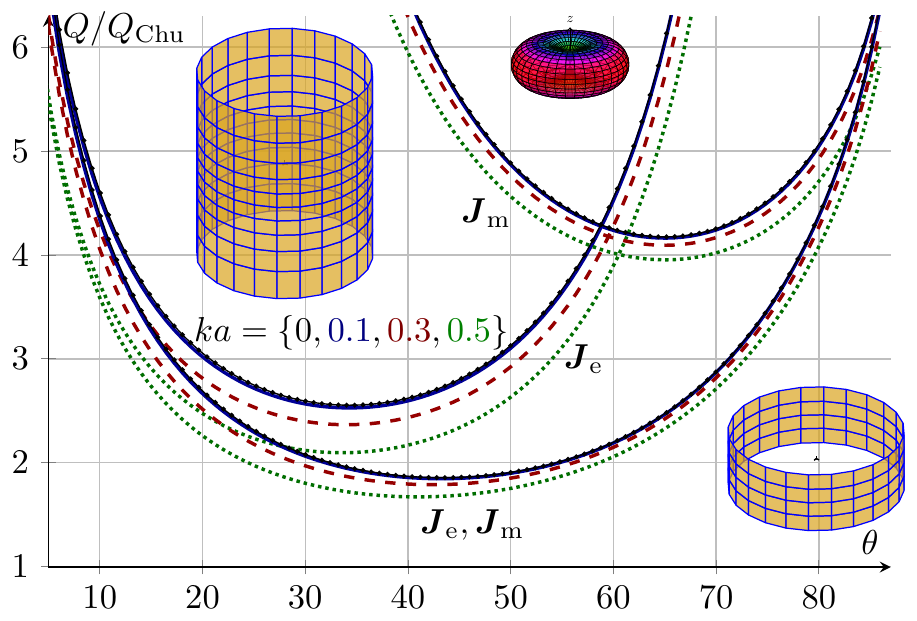}%
\hspace{3mm}
\includegraphics[width=0.45\columnwidth]{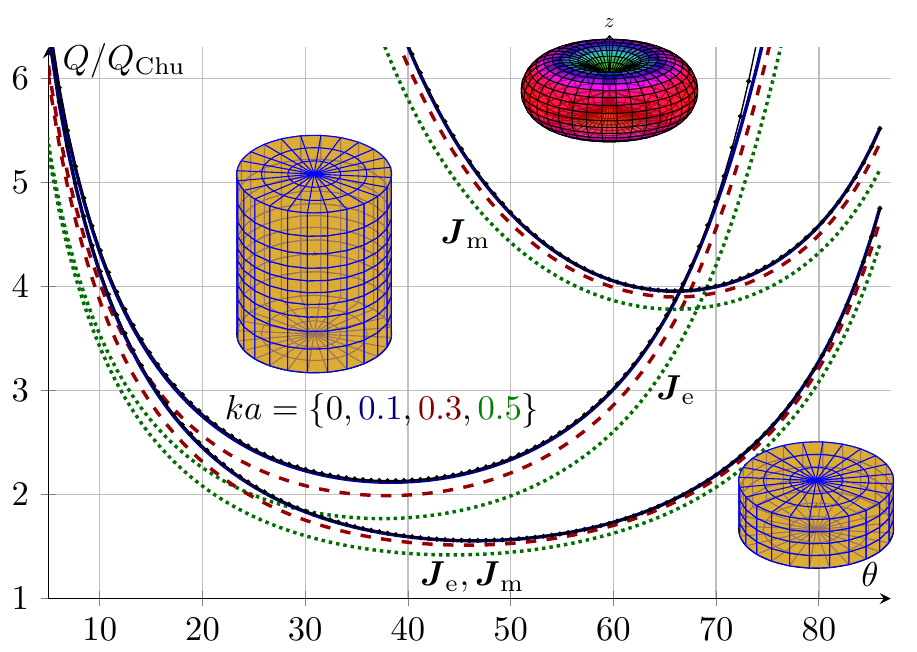}%
\caption{Minimum Q-factors for cylinder structures with aperture angle $\theta$ and radiating as an electrical dipole with electric currents, magnetic currents, and combined electric and magnetic currents. The resulting Q-factors are normalized with $k^3a^3$ for the cases $ka=\{0,0.1,0.3,0.5\}$. Antenna structure and its convex hull in the left and right figures, respectively.}%
\label{fig:Cylinder}%
\end{figure}

Cylindrical regions with height $2a\cos\theta$ and diameter $2a\sin\theta$ are considered to illustrate the minimum Q-factor~\eqref{eq:minQ} for structures that radiate as a $\zvh$ oriented electrical dipole in Fig.~\ref{fig:Cylinder}. The open cylinder consists of the curved surface. The top and bottom circular regions are added in the closed cylinder. The resulting Q-factors are depicted in Fig.~\ref{fig:Cylinder}. The Q-factors are normalized with $(ka)^{-3}$ and $Q_{\mrm{Chu}}=(ka)^{-3}+(ka)^{-1}$ in the top and bottom rows, respectively. 

The Q-factor is lowest for the combination of electric and magnetic current. The Q-factors have minimums around $\theta=\{30^{\circ},65^{\circ},45^{\circ}\}$ for the cases with electric, magnetic, and combinations of electric and magnetic currents, respectively. The lower Q-factor for electric currents for small polar angles $\theta$ is explained by the higher polarizability in the vertical than the horizontal directions for these thin elongated cylinders. The polarizability also explains the lower Q-factors for the magnetic currents for larger polar angles $\theta$ as the polarizability is higher in the horizontal directions for thick planar cylinders. 

The Q-factors decrease when the top and bottom is added to the open cylinder as seen in the right column in Fig.~\ref{fig:Cylinder}. This decrease is understood from the added degrees of freedom for the currents and also from the increased polarizability. The differences between the open and closed cylinders also confirm that the open structures have stored energy inside their convex hull. 

The dependence of the electrical size is illustrated with the $ka=\{0,0.1,0.3,0.5\}$ curves in Fig.~\ref{fig:Cylinder}. The major dependence of the electric size is removed by the normalization with $(ka)^{-3}$ and $Q_{\mrm{Chu}}=(ka)^{-3}+(ka)^{-1}$ in the top and bottom rows, respectively. The results confirm that the case with electric currents is well described by the $(ka)^{-3}$ normalization as shown by the forward scattering bound~\cite{Gustafsson+etal2009a}. The magnetic current case has a more complex dependence of the electrical size and is slightly better modeled with $Q_{\mrm{Chu}}=(ka)^{-3}+(ka)^{-1}$.   

\subsection{Open spherical shell}\label{S:ExSphShell}
\begin{figure}[t]%
\centering
\includegraphics[width=0.45\columnwidth]{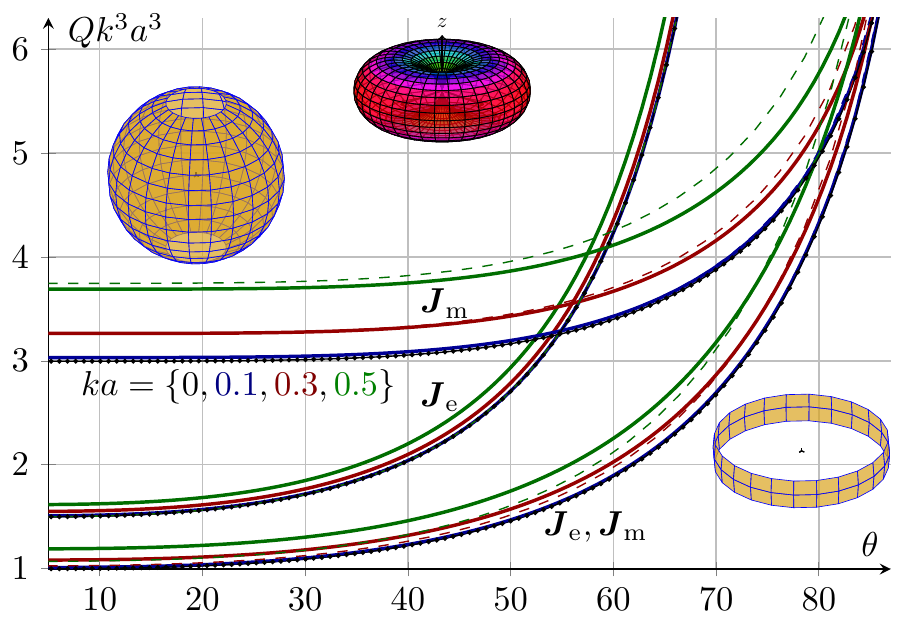}%
\hspace{3mm}
\includegraphics[width=0.45\columnwidth]{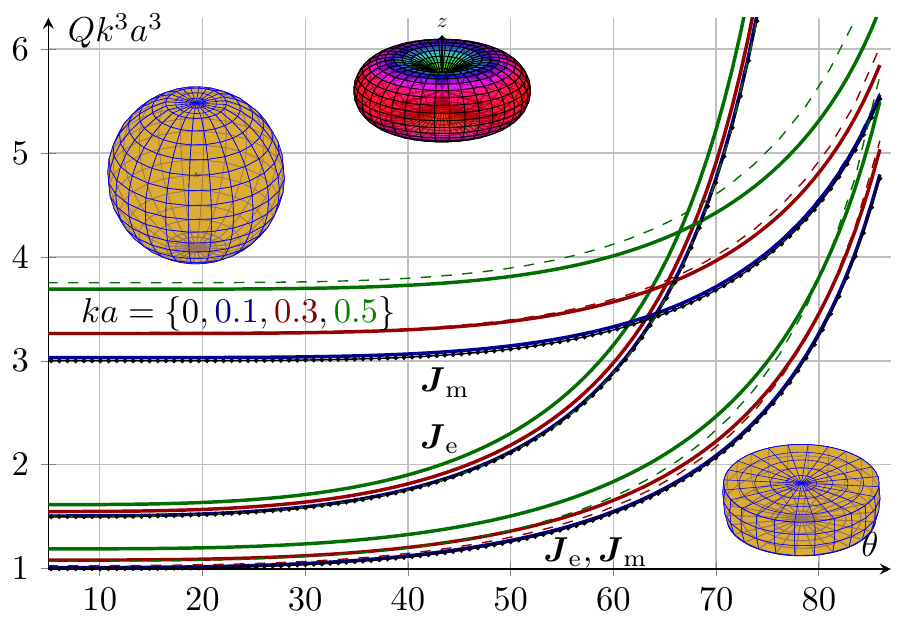}%
\caption{Minimum Q-factors for a spherical shell and its convex hull with aperture angle $\theta$ radiating as an electrical dipole with electric currents, magnetic currents, and combined electric and magnetic currents in solid curves. The minimum Q-factors are compared with the resulting Q-factors from the polarizability in the dashed curves, where $\tilde{Q}_{\mrm{m},0}=Q_{\mrm{m},0}(ka)^3Q_{\mrm{Chu}}$  and $Q_{\mrm{em},0}=(Q_{\mrm{e},0}^{-1}+\tilde{Q}_{\mrm{m},0}^{-1})^{-1}$ is used for the combined case.
The Q-factors are normalized with $(ka)^{-3}$ for the sizes $ka=\{0,0.1,0.3,0.5\}$. }%
\label{fig:SphereShell}%
\end{figure}

Open spherical shells with different opening angles, electrical sizes, and current excitations are used to illustrate minimum Q-factors. The convex optimization problem~\eqref{eq:minQ} to minimize the Q-factor for an antenna radiating as an $\hz$-directed electric dipole is solved with either electric currents, magnetic currents, and combination of electric and magnetic currents. The resulting Q-factors normalized with $k^3a^3$ are depicted in Fig.~\ref{fig:SphereShell}. The classical Chu~\cite{Chu1948} and Thal~\cite{Thal2006} bounds are retrieved for $\theta=0$. Is it observed that the combination of electric and magnetic currents has the lowest $Q$. This is obvious as the two other cases are special cases with one of the currents set to zero. The Q-factors also increase with increasing opening angle $\theta$ as this corresponds to having vanishing currents on parts of the sphere. The Q-factor for the electric case $\Jve$ is lower than for the $\Jvm$ case for small $\theta$ but increases faster and passes the magnetic case at $\theta\approx 55^{\circ}$. The minimum Q-factors are also compared with the resulting Q-factors from the electric and magnetic polarizabilities~\eqref{eq:Qka20}, where $Q_{\mrm{m},0}$ is normalized with $Q_{\mrm{Chu}}$, \ie $\tilde{Q}_{\mrm{m},0} = Q_{\mrm{m},0}(ka)^3Q_{\mrm{Chu}}$. The dashed lines corresponding to $Q_{\mrm{e},0}$, $\tilde{Q}_{\mrm{m},0}$ and $(Q_{\mrm{e},0}^{-1}+\tilde{Q}_{\mrm{m},0})^{-1}$ describes well the $Q$-factor obtained from the convex minimization problem~\eqref{eq:minQ}. 

%

\section{Conclusion}

We have derived expressions for the stored energies and radiated power for shapes that support both electric and magnetic sources.  The derivation is explicit and complete, with the aim to make the derivation easy to follow and verify.  
The derived stored energy expressions are based on the classic far-field subtracted stored energies~\eqref{WEM}, but as a difference to these energies they are both coordinate independent and non-negative.  
They are represented as the positive part of quadratic forms using a current-density source representation.  
It is observed that the singular kernels for the radiated power and stored energies can be described in terms of the EFIE and MFIE kernels, and non-singular terms are of a similar type.  The expressions are hence comparatively simple to implement and to test since they reduce to impedance like matrices.  
A detailed study of the electrically small case is given in~\cite{Jonsson+Gustafsson2015}.

The current-density representation of the far-field subtracted stored electric (magnetic) energy naturally consist of a core which is coordinate-independent and a rest-term with both coordinate-dependent and independent parts. The coordinate-independent core is easily identified by an inspection of the current-density representation of the far-field subtracted stored electric (magnetic) energy.  
With this current-density representation, we re-visit Chu's result.  The agreement between the coordinate independent part of the far-field subtracted stored energies and Chu's result is perfect, once we account for the subtraction of the far-field inside the sphere, \ie a difference of $ka$~\cite{Gustafsson+Jonsson2015b}.  
One discovery presented here, that the current-source representation makes explicit, is that the derived rest-term~\eqref{rest} of the stored energies contribute with a coordinate-independent contribution to the stored energies \eqref{Wemt} used to obtain the Q factor.  This is surprising since $\Wem^{\text{rest}}$ appears to be coordinate-dependent.  However, we conclude that in general $\Wem^\text{rest}$ contains a coordinate-dependent and a non-zero coordinate-independent part, as is illustrated for the sphere in Sec.~\ref{Coord}.  The coordinate-independent part of this term contributes and can lower the bounds for the stored energies, \eg using $\We$, $\Wm$. However, if we separate different orders of $ka$ we see that the rest term $\Wem^\text{rest}$ is a higher order correction term to the core-terms of the stored energies. 

This definition of stored energies is independent of the feed of the antenna. Thus we do not need to make any assumption on the feed \ie whether the feed is a current or a voltage source, nor any assumptions on how the feeding currents depend on frequency. The here used approach utilizes all possible currents in the antenna region to find the minimum antenna Q. Constraints on feeding can only increase the minimum antenna Q. We hence claim that the lower bounds on antenna Q derived here is a more general approach, than methods that require a particular feeding structure. 

Another feature of the here presented stored energies is that they are quadratic forms. They are thus suitable for a convex optimization problems. That is, given the support region of the current densities, open or closed, we illustrate how to determine the optimal antenna Q and its associated currents such that the structure radiates as an electric dipole. Several examples of such optimization for bodies of revolution are given. We can also show that the electric, magnetic, and the combined cases scale slightly different as a function of $ka$. Note that this type of optimization does not require or assume any a-priori requirement that the field vanish `inside' any particular region. Indeed, our examples above include current-surfaces which do not have an `inside'. The stored energies agree with the values of antenna Q for a sphere, but are applicable to arbitrary geometries as illustrated above.

\section{Acknowledgment}
We gratefully acknowledge the support from The Swedish Foundation for Strategic Research project AM13-0011 and from the Swedish Governmental Agency for Innovation Systems through the CHASE project NGAA. The support of UCN@Sophia LabEx for LJ's visit in April 2016 to the University of Nice Sophia Antipolis, France is gratefully acknowledged and a special thanks to Prof. F. Ferrero for his kind hospitality.

\appendix
\section{Kernels to far-field subtracted stored energies}\label{A}

The kernel $g$ in~\eqref{dF}--\eqref{dp} is an essential an essential part of the derivation of the far-field subtracted stored energies. In this appendix we reduce it from a volume integral to a sum of trigonometric type-functions. Towards this end recall~\eqref{g}:
\begin{equation}\label{w0}
g(\vr_1,\vr_2) = \lim_{r_0\rightarrow \infty}\int_{\br}  G_1G_2^* - \frac{\lexp{\ju k \hr\cdot (\vr_1-\vr_2)}}{16\pi r^2}\diff V,
\end{equation}
where $\br$ is a ball of radius $r_0$.  Here $G_1$ is defined by the outward propagating solution of $(\Laplace+k^2)G_1=-\delta(\vr-\vr_1)$ and similarly for $G_2$.  Differentiation with respect to $k$ yields $(\Laplace + k^2)\partial_k G_2 + 2kG_2=0$. We can thus write  
\begin{equation}\label{w1}
2k G_1G_2^* = -G_1(\Laplace+k^2)(\partial_k G_2^*) = 
-(\partial_k G_2^*)(\Laplace + k^2)G_1 - \Div\big(G_1\nabla \partial_k G_2^*-(\partial_k G_2^*)\nabla G_1\big),
\end{equation}
using standard vector identities for $\Div(\ldots)$.
Starting instead with $\partial_k G_1$ we find similarly that 
\begin{equation}\label{w2}
2k G_1G_2^* = 
-(\partial_k G_1)(\Laplace + k^2)G_2^* - \Div(G_2^*\nabla \partial_k G_1-(\partial_k G_1)\nabla G_2^*).
\end{equation}

Adding \eqref{w1} and \eqref{w2} results in $4kG_1G_2^*=\delta(\vr-\vr_1)\partial_k G_2^* + \delta(\vr-\vr_2)\partial_k G_1-\Div(\ldots)$. We hence have an identity for $G_1G_2^*$ in terms of a divergence term, and a term that is easy to integrate. Note that it is also `symmetric' with respect to $\vr_1$ and $\vr_2$. We integrate over $\br$, and recall that $\vr_1$ and $\vr_2$ are the position for the sources, \eg they are within $V\subset\br$ for $r_0>a$. Using Gauss theorem reduce the divergence term to a surface integral. We find for $r_0>a$ that
\begin{equation}
4k \int_\br G_1G_2^* \diff V = \partial_{k}(G_{12}^*+G_{12}) + \int_{\partial\br} g_2(\vr,\vr_1,\vr_2)\diff S.
\end{equation}
Since we determine the integral \eqref{w0} under the limit $r_0\rightarrow \infty$ we can without loss of generality assume that the radius of the surface $r_0$ is arbitrary large. To get the correct result we need only to keep terms up to order $r_0^{-2}$ in the integrand $g_2$. Thus utilizing that $r_0=|\vr|=r$ is large enough we determine the leading order behavior of the Green's function utilizing that $\vr\in \partial \br$:
\begin{multline}\label{divG}
g_2(\vr,\vr_1,\vr_2)=-\hr\cdot \Big(G_1\nabla \partial_k G_2^*-(\partial_k G_2^*)\nabla G_1
+G_2^*\nabla \partial_k G_1-(\partial_k G_1)\nabla G_2^*\Big)
\\= \frac{1}{16\pi^2}\hr\cdot \Big(k
\big(\frac{\hR_2+\hR_1}{R_1} + \frac{\hR_1+\hR_2}{R_2}\big) - \ju\frac{\hR_1}{R_1^2} +  \ju\frac{\hR_2}{R_2^2}
\Big)\lexp{-\ju k (|\vr-\vr_1|-|\vr-\vr_2|)} \\ \rightarrow  
 \frac{1}{16\pi^2}\Big(
\frac{4k}{r} + \frac{2k}{r^2}\hr\cdot(\vr_1 + \vr_2) - \frac{2k^2\ju }{r^2}\big(r_1^2-r_2^2-(\hr\cdot \vr_1)^2+(\hr\cdot \vr_2)^2\big)+ \mathcal{O}(\frac{1}{r^3})
\Big)\lexp{\ju k \hr\cdot (\vr_1- \vr_2)} .
\end{multline}
We note that the first term $k\exp{\ju k \hr\cdot(\vr_1-\vr_2)}/(4\pi^2 r)$ of $g_2$ in $\int_{\partial \br} g_2\diff S$ can be converted to a volume integral since we are integrating over a sphere. Thus $\int_0^r \diff r = r$ resulting in that $\int_{\br}  \frac{1}{r^2}f(\theta,\varphi)\diff V=r_0 \int_{\partial \Omega} f(\theta,\varphi)\diff \Omega$, where $\diff \Omega=\sin\theta\diff \theta\diff \varphi$ is the surface element on a unit sphere. We thus find:
\begin{multline}\label{gn}
\frac{-1}{4k}\int_{\partial \br} \big(G_1\nabla \partial_k G_2^*-(\partial_k G_2^*)\nabla G_1+
G_2^*\nabla \partial_k G_1-(\partial_k G_1)\nabla G_2^*\big) \cdot\hr \diff S
\\ = \int_{\br} \frac{\lexp{\ju k\hr\cdot(\vr_1-\vr_2)}}{16\pi^2 r^2} \diff V 
+ \frac{1}{32\pi^2} 
\int_{\Omega}  \lexp{\ju k \hr \cdot (\vr_1-\vr_2)} \hr\cdot(\vr_1+\vr_2)\diff \Omega  \\- \frac{\ju k }{32\pi^2} \int_{\Omega} 
\lexp{\ju k\hr\cdot(\vr_1-\vr_2)}(r_1^2 - r_2^2 - (\hr\cdot\vr_1)^2+(\hr\cdot\vr_2)^2)\diff \Omega 
+ \Oh(\frac{1}{r_0}).
\end{multline}
Above we have used $\Omega$ to denoted the unit sphere. 
The integrals on the last line are elementary but we write down for completeness. Let $\vp\in \RR^3$ be an arbitrary constant with length $p=|\vp|$ and direction $\hp=\vp/p$, then: 
\begin{equation}\label{int0}
\frac{1}{4\pi}\int_{\Omega} \lexp{\ju \hr\cdot \vp}\diff \Omega = \frac{\sin(p)}{p} = \sbj{0}(p).
\end{equation}
Here $\sbj{n}(p)$ is the spherical Bessel function of order $n$. 
The next integral follows directly from applying $-\ju \nabla_p$ upon~\eqref{int0}, we find 
\begin{equation}\label{int1}
-\ju \nabla_p \frac{1}{4\pi}\int_{\Omega} \lexp{\ju \hr\cdot \vp}\diff \Omega = 
\frac{1}{4\pi}\int_\Omega \hr \lexp{\ju \hr\cdot \vp}\diff \Omega = \ju \hp \frac{\sin p-p\cos p}{p^2} = \ju \hp\sbj{1}(p).
\end{equation}
The corresponding dyadic term integrand $\hr\hr\lexp{\ju \hr\cdot \vp}$ is determined by applying $\partial_{p_k}$ to~\eqref{int1} for $k=1,2,3$, to find
\begin{multline}\label{int2}
\frac{1}{4\pi}\int_{\Omega} \hr\hr \lexp{\ju \hr\cdot \vp}\diff \Omega = 
\frac{\sin p-p\cos p}{p^3} 
\boldsymbol{I} +
\big((p^2-3)\sin p + 3p\cos p\big)
\frac{\hp\hp}{p^3}
\\ = 
\frac{\sbj{1}(p)}{p}
\boldsymbol{I}
+\big(p\sbj{0}(p)-3\sbj{1}(p)\big)
\frac{\hp\hp}{p}
=
\frac{\sbj{1}(p)}{p}
\boldsymbol{I}
+\sbj{2}(p)
\hp\hp ,
\end{multline}
where $\boldsymbol{I}$ is the $3\times 3$ unit matrix.
To determine the integrals in~\eqref{gn} we let $\vp=k\vR_{12}=k(\vr_1-\vr_2)$ in the three elementary integrals~\eqref{int0}--\eqref{int2} above. Let $R_{12}=|\vR_{12}|$ and $\hR_{12}=\vR_{12}/R_{12}$. We also need to take the scalar products between $\vr_1+\vr_2$ and $\eqref{int1}$ and similarly scalar products to form $(\vr_k\cdot\hr)^2$ with $k=1,2$ for \eqref{int2}. Taking the limit that the radius $r_0\rightarrow \infty$ and collecting the terms give:
\begin{multline}\label{ggg}
g(\vr_1,\vr_2)  =\int_{\Rr}  G_1G_2^* - \frac{1}{16\pi^2} \frac{\lexp{\ju k \hr\cdot (\vr_1-\vr_2)}}{r^2}\diff V =  
-\frac{\sin(kR_{12})}{8 k \pi} -  \ju \frac{r_1^2-r_2^2}{8\pi k^2R_{12}^3}(\sin(kR_{12})-kR_{12}\cos(k R_{12}))  \\ =
-\frac{\sin(kR_{12})}{8 k \pi} -  \ju \frac{r_1^2-r_2^2}{8\pi R_{12}}\sbj{1}(kR_{12}) . 
\end{multline}
Note that~\eqref{ggg} differs from the expression derived in \cite{Vandenbosch2011}, where only the first term appear. To verify that the above calculation is correct we have also implemented it in Mathematica and the left and right hand side agree to numerical precision.

A similar identity  for the far-field reduced kernel concerns the integral:
\begin{equation}
\nabla_2 g(\vr_1,\vr_2) =  \nabla_2 \lim_{r_0\rightarrow\infty} \int_\br G_1G_2^* - \frac{\lexp{\ju k \hr\cdot(\vr_1-\vr_2)}}{(4\pi r)^2} \diff V. 
\end{equation}
It follows directly from~\eqref{ggg} that
\begin{equation}\label{nggg}
\nabla_2 g(\vr_1,\vr_2)=  \frac{1}{8\pi}\Big(\hR_{12}\cos(kR_{12}) + \ju (\frac{\vr_2+\vr_1}{R_{12}}-\hR_{12})\sbj{1}(kR_{12}) - \ju k\hR_{12} \frac{r_1^2-r_2^2}{R_{12}}\sbj{2}(kR_{12}) \Big). 
\end{equation}

\begingroup
\sloppy
\printbibliography
\endgroup

\end{document}